\documentclass[12pt,a4paper]{article}

\usepackage{latexsym}
\usepackage{amsmath}
\usepackage{amsfonts}
\usepackage{amssymb}
\usepackage{amsthm}
\usepackage{amscd}
\usepackage{mathrsfs}

\newcommand{\be}{\begin{equation}}
\newcommand{\ee}{\end{equation}}
\newcommand{\bea}{\begin{eqnarray}}
\newcommand{\eea}{\end{eqnarray}}

\renewcommand{\theequation}{\arabic{section}.\arabic{equation}}

\def\cA{{\cal A}}                       %
\def\cB{{\cal B}}
\def\cL{{\cal L}}                       %
\def\dt{\delta}                         %
\def\sing{\mathrm{sing}}                %
\def\cJ{{\cal J}}                       %
\def\bT{{\mathbb T}}                    %
\def\omfs{ \omega_{\mathrm{FS}}}        %
\def\cE{{\cal E}}                       %
\def\IIIb{\mathrm{III}_\mathrm{b}}      %
\def\cp{\mathbb{C}P^{n-1}}              %
\def\loc{\mathrm{loc}}                  %
\def\cR{{\cal R}}                       %
\def\reg{\mathrm{reg}}                  %
\def\can{\mathrm{can}}                  %
\def\cG{{\mathfrak{g}}}                 %
\def\ri{{\mathrm{i}}}                   %
\def\1{{\mbox{\boldmath $1$}}}          %
\def\red{\mathrm{red}}                  %
\def\bC{{\mathbb C}}                    %
\def\bR{{\mathbb R}}                    %
\def\bZ{{\mathbb Z}}                    %
\def\tr{\mathrm{tr}}                    %
\def\diag{\mathrm{diag}}                %
\def\SU{\mathrm{SU}}                    %
\def\SL{\mathrm{SL}}                    %
\def\su{\mathrm{su}}                    %
\def\cC{{\cal C}}                       

\begin{document}

\vspace*{0.5cm}
\begin{center}
{\Large \bf  New compact forms of  the trigonometric Ruijsenaars-Schneider system}
\end{center}

\vspace{0.2cm}

\begin{center}
L. Feh\'er${}^{a}$ and T.J. Kluck${}^b$  \\

\bigskip

${}^a$Department of Theoretical Physics, University of Szeged\\
Tisza Lajos krt 84-86, 6720 Szeged, Hungary, and\\
Department of Theoretical Physics, WIGNER RCP, RMKI\\
1525 Budapest, P.O.B.~49,  Hungary \\
e-mail: lfeher@physx.u-szeged.hu

\bigskip

${}^b$Mathematical Institute, Utrecht University\\
P.O.\ Box 80010, 3508 TA Utrecht, the Netherlands\\
e-mail: tkluck@infty.nl

\bigskip

\end{center}

\vspace{0.2cm}

\begin{abstract}
The  reduction of the quasi-Hamiltonian double of $\SU(n)$ that has been shown to underlie
 Ruijsenaars' compactified  trigonometric $n$-body system is
studied in its natural generality. The constraints contain a parameter $y$,
restricted in previous works  to $0<y < \pi/n$
because Ruijsenaars' original compactification relies on an equivalent condition.
It is found that  allowing generic $0<y<\pi/2$
 results in the appearance of new
self-dual compact forms, of two  qualitatively different types
depending on the value of $y$.
The type (i) cases are similar to the standard case in that the reduced phase space
comes equipped with globally smooth action and position variables, and turns out to be
symplectomorphic to $\cp$ as a Hamiltonian toric manifold.
In the type (ii) cases  both the position variables and the action variables
develop singularities on a nowhere dense subset.
A full classification is derived for the  parameter $y$ according to the
type (i) versus type (ii) dichotomy.
The simplest new type (i) systems, for which  $\pi/n < y < \pi/(n-1)$,
are described in some detail as an illustration.

\end{abstract}

\newpage

\section{Introduction}
\setcounter{equation}{0}

The integrable many-body systems discovered by Ruijsenaars and Schneider \cite{RS}
are popular due to their rich mathematical structure and connections to
important areas of physics.
These systems appear in
topics extending from soliton equations to gauge theories and  representation theory
(see e.g.~\cite{RS,SR-Kup, GN, FR, JHEP,GK,  EK, Ob}).
As limiting cases they contain the non-relativistic Calogero-Moser systems
that also have many applications \cite{SR-CRM,Sut,Eti}.
Several members of this family have been realized as Hamiltonian reductions of
higher dimensional ``free systems''  (\cite{KKS,OPI,FKinCMP,Puszt,Mars} and references therein), which
permits an understanding of their dynamics and duality  properties \cite{SR-CMP,RIMS95} in group-theoretic terms.
In the current work  \emph{new} variants of the Ruijsenaars-Schneider (RS) system  will be derived
by exploiting the reduction method.

This paper is a continuation of joint work of the first author with
Klim\v c\'\i k \cite{FK}, where the self-dual compactified trigonometric RS system
of Ruijsenaars \cite{RIMS95}  was interpreted as  a reduced system arising from a double of $G:=\SU(n)$.
A key point of the quasi-Hamiltonian reduction used in \cite{FK} was the fixing of
the $G$-valued moment map to the maximally degenerate non-scalar matrix
\be
\mu_0(y):=\diag\left(e^{2\ri y}, \ldots, e^{2\ri y}, e^{-2(n-1) \ri y}\right)
\label{I1}\ee
with
\be
0<\vert y \vert < \pi/n.
\label{I2}\ee
The restriction (\ref{I2}) on the angle-parameter $y$ was adopted in \cite{FK} from the very beginning,
motivated (solely) by its eventual identification with a corresponding parameter in the
``$\IIIb$-system''
of Ruijsenaars \cite{RIMS95}, where it was restricted to this range based on intuitive arguments.

The observation that prompted the present work is that in the scheme of quasi-Hamiltonian reduction there is
no internal reason that requires restriction of the parameter $y$ to the above range.
Our goal is to explain that for any
generic\footnote{As in the Abstract, one may restrict
to $0<y<\pi/2$ without losing generality.}
$y\in (-\pi/2, \pi/2)$, the reduction built on the moment map value $\mu_0(y)$  always leads to a compact version
of the trigonometric RS system, which is
 not equivalent to the one constructed in
\cite{RIMS95} unless (\ref{I2}) holds.

Before turning to the content of this paper, we  need to recall some essential
points of \cite{FK}.
The starting point there is the so-called internally fused double \cite{AMM} of $G$,
given by
\be
G \times G  = \{(A,B)\}
\label{I3}\ee
 equipped with the 2-form
\be
\omega^\lambda := \lambda \left( \langle A^{-1} \mathrm{d}A \stackrel{\wedge}{,} \mathrm{d}B B^{-1}\rangle
+\langle  \mathrm{d}A A^{-1} \stackrel{\wedge}{,} B^{-1} \mathrm{d}B \rangle
- \langle (AB)^{-1} \mathrm{d} (A B) \stackrel{\wedge}{,} (BA)^{-1} \mathrm{d} (BA)
\rangle \right),
\label{I4}\ee
where $\lambda\neq 0$ is an arbitrary real constant and
$\langle X, Y\rangle := - \frac{1}{2} \tr(XY)$.
The $2$-form,  the moment map
\be
\mu\colon (A,B) \mapsto AB A^{-1} B^{-1},
\ee and
the componentwise conjugation action of $G$ on $G\times G$, whereby
\be
 G \times (G\times G) \ni (\eta, (A,B)) \mapsto
(\eta A \eta^{-1}, \eta B \eta^{-1})\in G\times G,
\label{I5}\ee
satisfy the axioms of a quasi-Hamiltonian space \cite{AMM}.
As a result, the reduced phase space
\be
P(\mu_0): = \mu^{-1}(\mu_0)/G_{\mu_0}
\label{I6}\ee
becomes (whenever it is smooth) a \emph{symplectic} manifold.
By applying the smooth class functions of $G$ to either components of the pair
$(A,B)\in G\times G$, one obtains two sets of $G$-invariant functions on $G\times G$ that descend to two
Abelian Poisson algebras on $P(\mu_0)$.
Therefore $(n-1)$ independent class functions of $G=\SU(n)$ may reduce to
Liouville integrable Hamiltonian systems if $P(\mu_0)$ is a smooth manifold of dimension $2(n-1)$.
Note that $P(\mu_0)$ is always compact and connected \cite{AMM} and
the choice of $\mu_0$  matters only up to conjugation.
It turns out that
the dimension of $P(\mu_0)$ is $2(n-1)$ if $\mu_0$ is conjugate to
$\mu_0(y)$ of the form (\ref{I1}) with generic
$y\in (-\pi/2, \pi/2)$.

Under the restriction (\ref{I2}),
the reduced phase space  was identified in \cite{FK} as
the complex projective space $\cp$ carrying  a multiple of the standard
Fubini-Study symplectic form.
The analysis relied on
the one-to-one parametrization of the conjugacy classes of $\SU(n)$ by the Weyl alcove
\be
\cA:= \{\xi \in \bR^n\,\vert\, \xi_k \geq 0\quad (\forall k=1,\ldots, n), \quad \xi_1 + \cdots + \xi_n = \pi\}.
\label{I7}\ee
Concretely, $\xi\in \cA$ labels the conjugacy class represented by the diagonal
matrix
\be
\delta(\xi) = \diag(\delta_1(\xi), \ldots, \delta_n(\xi)),
\quad
\delta_1(\xi):=e^{\frac{2\ri}{n}\sum_{j=1}^n j\xi_j}
\quad
\delta_{k+1}(\xi):= e^{2\ri \xi_k} \delta_k(\xi).
\label{I8}\ee
In order to present the characterization of the reduced system,
introduce the ``Weyl alcove with thick walls''
\be
\cA_{ y}:
= \{ \xi \in \cA \,\vert\, \xi_k \geq \vert y \vert\ \quad(\forall k=1,\ldots, n)\},
\quad \hbox{for any}\quad
0 <\vert y \vert < \pi/n,
\label{I10}\ee
and let $\cA^+_{ y  }$ be the interior of $\cA_{ y }$.
Consider the torus $\bT^{n-1}$ with elements
$(e^{\ri \theta_1}, \ldots, e^{\ri \theta_{n-1}}) \in \bT^{n-1}$
and
equip the Cartesian product $\cA_y^+ \times \bT^{n-1}$ with the symplectic form
\be
\Omega^\lambda_\can := \lambda \sum_{k=1}^{n-1} \mathrm{d}\theta_k \wedge \mathrm{d} \xi_k.
\label{I9}\ee
Finally, extend the above definitions by the  convention
\be
\delta_{k+n}:= \delta_{k},
\quad
\xi_{k+n}:= \xi_{k},
\quad
\theta_{k+n}:= \theta_k \quad\hbox{and}\quad \theta_0:=0.
\label{I11}\ee

In \cite{FK} a dense open submanifold of the reduced phase space $P(\mu_0(y))$ was exhibited
which is symplectomorphic to $(\cA^+_{ y } \times \bT^{n-1}, \Omega^\lambda_\can)$ and permits
the identification
of the reduction of the invariant function $\Re(\tr(A))$ as the local Hamiltonian
\be
H_{ y }^\loc (\xi, \theta) := \sum_{j=1}^n \cos(\theta_j - \theta_{j-1})
\prod_{k=j+1}^{j+n-1}
\left\vert 1 +4 \frac{\sin^2 y}{ \left[(\delta_k/\delta_j)^{1/2}  -
 (\delta_k/\delta_j)^{-1/2}\right]^2 }  \right\vert^\frac{1}{2}.
\label{I12}\ee
Here, the square roots $(\delta_k/\delta_j)^{1/2}$ are a notational convenience, and we
do not actually pick a branch for the square root since the square roots formally disappear
after expanding the square.
This Hamiltonian can be interpreted in terms of the interaction
of $n$ ``particles'' on the unit circle, located at $\delta_1,\ldots, \delta_n$.
Using (\ref{I11}), one has
\be
\delta_k = \delta_j e^{2\ri (\xi_j + \cdots + \xi_{k-1})},
\quad
\forall k=j+1,\ldots, j+n-1,
\label{I13}\ee
and  the Hamiltonian takes the Ruijsenaars-Schneider form of $\IIIb$ type \cite{RIMS95}:
  \be
H_{ y }^\loc(\xi,\theta) =  \sum_{j=1}^n \cos(\theta_j - \theta_{j-1}) \prod_{k=j+1}^{j+n-1}
\left\vert 1 - \frac{\sin^2 y}{ \sin^2(\sum_{m=j}^{k-1} \xi_m)} \right\vert^\frac{1}{2}.
\label{I14}\ee
The condition $\xi_k \geq \vert y\vert$ in (\ref{I10}) means that the particles have a minimal
angular distance given by $2 \vert y\vert$, and this ensures that all functions under
 the absolute values above are non-negative.
Since $\xi_1 +\cdots +\xi_n  = \pi$,
these features can occur only for $\vert y\vert \leq \pi/n$.
In \cite{RIMS95} these features were deemed desirable, and hence
 $y$ was restricted to the range (\ref{I2}).

It is of course superfluous to write absolute values in the formulae (\ref{I12}) and (\ref{I14})
if all the relevant functions are non-negative.
Our usage anticipates that there exist \emph{new}  systems having perfectly
reasonable global properties and a similar local description as above,
with the difference that some factors under the absolute values
in the local formula (\ref{I14}) are non-positive.
In fact, we shall demonstrate that for generic
parameter $y$ from  the full range $(-\pi/2, \pi/2 )$ the quasi-Hamiltonian
reduction built on $\mu_0(y)$ (\ref{I1})
leads to a smooth  reduced phase space
that contains a maximal dense open submanifold parametrized by
$\cA_{ y }^+ \times \bT^{n-1}$, for some open $\cA_{ y }^+ \subset \cA$,
 on which the symplectic form is provided by $\Omega^\lambda_\can$ (\ref{I9})
 and the principal reduced Hamiltonian
$\Re(\tr(A))$ is given (in general up to a sign) by the formula (\ref{I14}).
In the general case, the domain $\cA^+_{ y}\subset \cA$  will be
identified as a certain dense open subset
of the set of $\xi \in \cA$ for which $\delta(\xi)$
represents the conjugacy class of some regular unitary matrix $B$ entering
a pair $(A,B) \in \mu^{-1}(\mu_0(y))$.
One of the main issues studied in the text  is
the dependence of $\cA^+_{y}$ on $y$.

We shall classify the coupling parameter $y$ according to the criterion of whether the relation
\be
\mu^{-1}(\mu_0(y)) \subset G_\reg \times G_\reg
\label{T2}\ee
is valid or not, i.e., whether it is true or not that
the constraint surface contains only regular matrices.
The cases verifying (\ref{T2}) will be later called \emph{type (i)} and those
that violate (\ref{T2}), \emph{type (ii)}. The relation (\ref{T2})
is known to hold in the standard case. Its validity
guarantees that the distinct eigenvalues of $A$ and $B$ descend to \emph{smooth} functions on
the reduced phase space and give rise to globally smooth
action variables and position variables of the associated compact RS system.
Said in more technical terms, if (\ref{T2}) holds, then the reduced system carries two
distinguished Hamiltonian torus action.

Our main new result is that we shall find all
$y$ values verifying (\ref{T2}), and shall prove that in these cases
the reduced phase space is symplectomorphic to $\cp$ with a multiple of the Fubini-Study
symplectic structure.
In fact, in these cases $\cA_y^+$ will turn out to be an open simplex, whose closure
lies in the interior of $\cA$ and yields the moment polytope of the corresponding torus action.
As listed by Theorem 12 in Section 3,
there are many new cases different from (\ref{I2}) which fall
into this category. The simplest such new cases are associated with the range
\be
\pi/n<\vert y \vert < \pi/(n-1), \qquad
n\geq 3,
\label{I15}\ee
for which we obtain that
\be
\cA^+_{ y } = \{ \xi\in \cA\,\vert\, \xi_k < \vert y \vert \quad(\forall k=1,\ldots, n)\}.
\label{I16}\ee
We shall describe these examples in some detail,
and show that the compact RS systems associated with the ranges
(\ref{I2}) and (\ref{I15})
represent non-equivalent many-body systems.
This means that the respective many-body Hamiltonians cannot be
converted into each other by a canonical transformation
that maps coordinates into coordinates.
The same conclusion can be reached regarding any two coupling parameters $y_1$ and $y_2$ for which
$\sin^2 y_1\neq \sin^2 y_2$.
We remark in passing that if $\xi$ belongs to the domain (\ref{I16}),  then
precisely two of the factors under the absolute values in (\ref{I14}) are negative
for each $j=1,\ldots, n$.

The globally smooth class functions of $G$ descend to smooth reduced Hamiltonians
in involution also in the cases for which (\ref{T2}) is not valid,  and engender
Liouville integrable systems.
However, the action variables and the position variables arising from the eigenvalues
of $A$ and $B$ develop singularities at the loci of the coinciding eigenvalues, which
intersect $\mu^{-1}(\mu_0(y))$ when (\ref{T2}) does not hold.
The actions and the positions enjoy a duality relation in all
our reduced systems, and thus their qualitative properties are the same.
This duality stems from
a natural $\SL(2,\bZ)$ symmetry between $A$ and $B$ in the pair $(A,B)\in G\times G$,
which survives  reduction for any moment map value $\mu_0\in G$ \cite{FK}.

As for the content of the rest of the paper,  we first note that
many of our arguments will be adaptations of arguments from \cite{FK}.
We do not wish to repeat those in detail,  but need to state clearly what
changes and what remains true if the restriction (\ref{I2}) is dropped.
This is done in Section 2, where we generalize relevant results from \cite{FK}.
This section contains also  significant novel results, e.g.,
the description of the
fixed points of the torus action given by Corollary 4 of Lemma 3 and the important Theorem 6.
Then we present entirely new results in Section 3.
 Theorem 7 gives the form of $\cA_y^+$ for any generic
$y$.  Theorems 12 and 13 describe the full set of
 type (i) cases, i.e., all cases satisfying (\ref{T2}).
As illustration, the simplest new systems  of type (i)
 are detailed in Section 4.
An example violating (\ref{T2}) will be also exposed briefly at the end of Section 4.
The results and open problems
are discussed in Section 5,
and certain non-trivial details are relegated to appendices.

In Sections 1 and 2 it is often assumed that
 $- \pi/2 < y < \pi/2$, while
in Section 3 it will be more convenient to speak in terms of
 $0 <y < \pi$. This should not cause any confusion, since
$y$ enters through $\mu_0(y)$ (\ref{I1}) and thus can matter at most modulo $\pi$.
It is also worth noting that componentwise complex conjugation
of the pair $(A,B)$ gives an anti-symplectic diffeomorphism between $P(\mu_0(y))$ and $P((\mu_0(y))^{-1})$.
By using this, it would be possible to restrict attention to $0<y < \pi/2$ without losing generality,
but  we here find it advantageous not to do so.

\section{Results for generic value of the coupling parameter}
\setcounter{equation}{0}

We are interested in those reductions for which
the reduced phase space (\ref{I6}) is a smooth manifold of dimension $2(n-1)$.
It is readily extracted from Subsection 3.1 of \cite{FK} that this holds if and only if
$e^{2\ri y}$ is not an $m$-th root of unity for any $m=1,2,\ldots, n$.
In these cases the isotropy group\footnote{Note from (\ref{I5}) that  only the group $G/\bZ_n=
\mathrm{ U}(n)/\mathrm{U}(1)$ acts effectively on the double. For notational convenience, we
will occasionally use the non-effective $\mathrm{U}(n)$-action instead.}
$G_{\mu_0(y)}/\bZ_n=\mathrm{ U}(n)_{\mu_0(y)}/\mathrm{U}(1) $
acts freely on $\mu^{-1}(\mu_0(y))$.
We henceforth assume that $y$ satisfies
\be
e^{2 \ri y m} \neq 1,
\qquad
\forall m=1,2,\ldots, n.
\label{T1}\ee
One of the important points explained below is that if the relation
$\mu^{-1}(\mu_0(y)) \subset G_\reg \times G_\reg$
(\ref{T2})
is valid, then the reduced phase space is a Hamiltonian
toric manifold.  This means that $P(\mu_0(y))$ carries the effective Hamiltonian action
of an $(n-1)$-dimensional torus $\bT^{n-1}$.
In other words, under (\ref{T2})  we obtain a compact
integrable Hamiltonian system having globally smooth action variables \cite{ACL}.
Independently if (\ref{T2}) holds or not, we shall prove
that the reduction leads to an integrable
system on $P(\mu_0(y))$, which contains a dense open submanifold where the
principal Hamiltonian descending from $\Re \tr(A)$ with $(A,B) \in \mu^{-1}(\mu_0(y))$
takes the RS form.

\subsection{Recall of the $\beta$-generated torus action}

Following \cite{FK}, let us define the ``spectral function'' $\Xi\colon G \to \cA$ by the requirements
\be
\Xi(\delta(\xi)):= \xi \quad\hbox{and}\quad
\Xi( \eta g \eta^{-1}) =\Xi(g),
\quad \forall \eta, g\in G.
\label{T3}\ee
Note that $\Xi$ is $G$-invariant, its real component functions are
globally continuous on $G$, and
their restrictions to $G_\reg$ belong to $C^\infty(G_\reg)$.
It is also important to know that $\Xi$  is not differentiable
at $G_\sing = G \setminus G_\reg$ consisting of matrices with
multiple eigenvalues (see Appendix A).
It follows that the functions
\be
\alpha\colon (A,B)\mapsto \Xi(A)
\quad\hbox{and}\quad
\beta\colon (A,B) \mapsto \Xi(B)
\label{T4}\ee
engender continuous maps
\be
\hat \alpha\colon P(\mu_0(y)) \to \cA
\quad\hbox{and}\quad
\hat \beta\colon P(\mu_0(y)) \to \cA.
\label{T5}\ee
These maps are globally smooth if (\ref{T2}) is valid, in which case they take
their values in the interior of the alcove, denoted
\be
\cA^\reg:= \{ \xi \in \cA\,\vert\, \xi_k \neq 0\quad (\forall k=1,\ldots, n)\,\}.
\label{Areg}\ee
Throughout this section, we restrict our attention to the open submanifold
\be
\hat \beta^{-1}(\cA^\reg) \subset P(\mu_0(y)),
\label{T12}\ee
where the components of $\hat \beta$ are $C^\infty$ functions.
This submanifold equals $P(\mu_0(y))$ if (\ref{T2}) holds, and it will be
shown to be an open dense subset for any $y$ satisfying (\ref{T1}).

The linearly independent smooth functions
\be
\hat \beta^\lambda_j := \lambda \hat \beta_j,
\quad
j=1,\ldots, n-1,
\label{T6}\ee
induce Hamiltonian flows on the submanifold (\ref{T12}).
These are $2\pi$-periodic
and thus generate a $\bT^{n-1}$-action \cite{FK}.
To describe this torus action, let us take a representative
 $(A,B) \in \mu^{-1}(\mu_0(y))$ of the point $[(A,B)] \in P(\mu_0(y))$.
Then diagonalize $B$, that is, introduce $\xi \in \cA^\reg$ and $g \in G$ by
\be
g B g^{-1} = \delta(\xi).
\label{T7}\ee
The action of
\be
\tau = (\tau_1,\ldots, \tau_{n-1})\in \bT^{n-1}
\label{T8}\ee
is provided by the following formula:
\be
\hat \Psi^\beta_\tau: [(A,B)] \mapsto [( A g^{-1} \varrho(\tau) g, B )]
\label{T9}\ee
with
\be
\varrho(\tau):= \diag(1/\tau_1, \tau_1/\tau_2, \tau_2/\tau_3,\ldots,
\tau_{n-2}/\tau_{n-1}, \tau_{n-1}).
\label{T10}\ee
It can be shown (see below) that this Hamiltonian $\bT^{n-1}$-action
on $\hat\beta^{-1}(\cA^\reg)$,
which we call the $\beta$-generated torus action\footnote{
Of course one also has an analogously operating $\alpha$-generated torus action
on $\hat\alpha^{-1}(\cA^\reg)$ \cite{FK}.},
is an effective action.

Since  $P(\mu_0(y))$ is compact and connected \cite{AMM}, we see that
\emph{$P(\mu_0(y))$ is a Hamiltonian toric manifold under the $\beta$-generated
torus action whenever (\ref{T2}) is valid}.
Then we can invoke the powerful Atiyah-Guillemin-Sternberg and Delzant
theorems of symplectic geometry \cite{De, Can} that determine the structure of a Hamiltonian
toric manifold in terms
of the moment map.
In particular, under (\ref{T2}), we know that the image of the map
$\hat \beta$ is a closed convex polytope in $\cA^\reg$. The polytope is
the convex hull of its vertices, which are the images of the
fixed points of the $\beta$-generated torus action. The correspondence
between the vertices and the fixed points is one-to-one.
Moreover, the polytope completely characterizes the Hamiltonian toric manifold.

On account of the above, at least in the presence of (\ref{T2}), we may establish the structure
of $P(\mu_0(y))$ if we can find its image under the map
$\hat \beta$.
Next, we shall present a characterization of the image
$\beta\left( \mu^{-1}(\mu_0(y)) \cap (G \times G_\reg)\right)$,
and study the equations that determine fixed points of the $\beta$-generated torus action.

\subsection{The $\beta$-regular part of the reduced phase space}

The open submanifold $\hat\beta^{-1}(\cA^\reg)$ (\ref{T12}) will be
called the $\beta$-regular part of the reduced phase space.
Here, we are interested in the $\hat{\beta}$-image of this submanifold, given by
\be
\cA_y^\reg:= \hat \beta(P(\mu_0(y)) \cap \cA^\reg =
\beta\left( \mu^{-1}(\mu_0(y)) \cap (G \times G_\reg)\right).
\label{T13}\ee
Our description of this image relies on
the functions
$z_\ell(\xi,y)$ defined on $\cA^\reg$  by the formula
\be
z_\ell(\xi,y):=
\frac{ e^{2\ri y}-1}{ e^{2n\ri y}-1}
\prod_{\substack{j=1\\j \neq \ell}}^{n}\frac{ \delta_j(\xi) -   e^{2\ri y} \delta_{\ell}(\xi)}
{\delta_j(\xi)-\delta_\ell(\xi)},
\qquad
\forall \ell=1,\ldots, n.
\label{T14}\ee
By using formula (\ref{I8}) and the periodicity convention (\ref{I11})  we can spell out
this function as
\be
z_\ell(\xi,y) =
      \frac{\sin(y)}{\sin(ny)}
      \prod^n_{\substack{j=1\\j \neq \ell}} \frac{e^{\ri y}\delta_\ell - e^{-\ri y}\delta_j}
                           {\delta_\ell - \delta_j}
= \frac{\sin(y)}{\sin(ny)}
\prod_{j=\ell+1}^{\ell+ n-1}
\left[  \frac{\sin(\sum_{m=\ell}^{j-1}\xi_m - y)}{\sin(\sum_{m=\ell}^{j-1} \xi_m)} \right].
\label{T15}\ee
The proof of the following result can be extracted from Section 3.2 of \cite{FK}.
Nevertheless we sketch  it here since it is required for our later arguments.

\medskip
\noindent
{\bf Lemma 1.} \emph{The element $\xi \in \cA^\reg$ belongs to the $\hat \beta$-image
(\ref{T13}) if and only if
$z_\ell(\xi,y)$ is non-negative for all $\ell=1,\ldots,n$.}

\medskip
\noindent
{\bf Proof.}
Suppose that we have
\be
AB A^{-1} B^{-1} = \mu_0(y).
\label{T16}\ee
Since $B$ is conjugate to $\delta(\xi)$ with some $\xi\in \cA$,  (\ref{T16}) is equivalent to
\be
A^g \delta(\xi) (A^g)^{-1} = (g \mu_0(y) g^{-1}) \delta(\xi),
\label{T17}\ee
where $g$ is a unitary matrix for which
\be
\delta(\xi) = g B g^{-1}
\quad\hbox{and}\quad A^g = g A g^{-1}.
\label{T18}\ee
Denoting by $v\in \bC^n$ the last column of the matrix $g$,
\be
v_\ell := g_{\ell n},
\label{T19}\ee
it is easily checked that
\be
g \mu_0(y) g^{-1}=
 e^{2\ri y}\1_n+ (e^{2\ri(1-n) y}-e^{2\ri y}) vv^\dagger.
 \label{T20}\ee
Equation (\ref{T17})  implies the equality of the characteristic polynomials of the matrices
on the two sides, which gives
\be
\prod_{j=1}^n (\dt_j(\xi) -x )=\prod_{j=1}^n (\dt_j(\xi) e^{2\ri y}- x)
+ (e^{2\ri(1-n) y}-e^{2\ri y}) \sum_{k=1}^n \Bigl(\vert v_k \vert^2 \dt_{k}(\xi)
\prod_{\substack{j=1\\j \neq k}}^n (\dt_j (\xi)e^{2\ri  y} -x)\Bigr)
\label{T21}\ee
for all $x \in \bC$.
Supposing now that $B$ is regular, evaluation of (\ref{T21}) at the $n$ distinct values
 $x=\delta_{\ell}(\xi)e^{2\ri y}$ leads to the equations
\be
\vert v_\ell\vert^2=z_\ell(\xi,y)
\label{T22}\ee
with the functions defined in (\ref{T14}). Therefore these functions must be non-negative
for all $\xi$ in the image (\ref{T13}).

Conversely, suppose that all $z_\ell$ in (\ref{T14}) are non-negative at $\xi \in \cA^\reg$.
Choose $v=v(\xi,y)\in \bC^n$ for which
$\vert v_\ell\vert^2 = z_\ell(\xi,y)$.
Then we observe that the equality (\ref{T21}) holds at all $x\in \bC$ since we can check that
it holds at the $n$ distinct values $\delta_\ell(\xi) e^{2\ri y}$.
Evaluating this equality at $x=0$ implies that the vector $v(\xi,y)$ has unit norm, and
consequently the right-hand-side of (\ref{T20}) with this vector
defines a unitary matrix of unit determinant, now denoted as $\mu_{v(\xi,y)}$.
Since (\ref{T21}) guarantees that the unitary matrices $\delta(\xi)$
and $\mu_{v(\xi,y)}\delta(\xi)$
have the same spectra, there  exists a unitary matrix, say $A_0$, for which
\be
A_0 \delta(\xi) A_0^{-1} = \mu_{v(\xi,y)} \delta(\xi),
\label{T23}\ee
and we can normalize $A_0$ to have unit determinant, yielding $A_0 \in G$.
Then we take a unitary matrix $g$ having $v(\xi,y)$ as its last column and
conjugate both sides of (\ref{T23}) by $g^{-1}$. This allows to conclude that
\be
A:= g^{-1} A_0 g
\quad\hbox{and}\quad
B:= g^{-1} \delta(\xi) g
\label{T24}\ee
satisfy (\ref{T16}), i.e., $(A,B) \in \mu^{-1}(\mu_0(y))$ and $\beta(A,B)=\xi$ holds.
 \qed

\medskip
\noindent
{\bf Remark 2.}
The special element $\xi^* \in \cA^\reg$ having equal components
\be
\xi^*_k:= \pi/n, \qquad
\forall k=1,\ldots, n,
\label{T25}\ee
is in the image (\ref{T14}) for all allowed values of $y$.
Indeed, one can check that
\be
z_\ell(\xi^*,y) = \frac{\sin(y)}{\sin(ny)}
\prod_{k=1}^{n-1}\left[  \frac{\sin(k \frac{\pi}{n} - y)}{\sin( k\frac{\pi}{n})} \right] >0,
\qquad \forall \ell=1,\ldots, n,
\label{T26}\ee
at any admissible value of $y$. The point is that
if
$ \frac{m\pi}{n} < y < \frac{(m+1)\pi}{n}$ for some $m=0,\ldots, n-1$, then  $m$ factors in
the above product are negative and $(n-m-1)$ factors are positive\footnote{We here
took $y$ from the interval $(0,\pi)$ instead of $(-\pi/2, \pi/2)$, which
is permitted since only its value modulo $\pi$ appears in $\mu_0(y)$ (\ref{I1}).} .
This yields exactly the right parity to cancel the possible minus sign from $\sin(ny)$.

\medskip
As a spin-off from the above proof, we can in principle
construct all elements of the $\beta$-regular part of the constraint surface $\mu^{-1}(\mu_0(y))$
by the following algorithm.
 First, take $\xi\in \cA^0$ for which $z_\ell(\xi,y)$ is non-negative for all $\ell$,
and define
\be
v_\ell(\xi,y):= \sqrt{z_\ell(\xi,y)}
\label{T27}\ee
using non-negative square roots.
Choosing a unitary matrix $g:= g(v)$ that has $v$ as its last column and taking $A_0\in G$
subject to (\ref{T23}), define $(A,B)$ according to (\ref{T24}).
Then the most general element of $\mu^{-1}(\mu_0(y))$ for which $\beta$ takes the value $\xi$
is a gauge transform of an element of the following form:
\be
(A g(v)^{-1} \varrho g(v) ,B)
\quad
\hbox{with}\quad \varrho \in \mathrm{S}\bT^n:= \mathrm{SU}(n) \cap \bT^n.
\label{T28}\ee
This holds because  equation (\ref{T23}) determines
$A_0$ up to right multiplication by a diagonal matrix, leading to $\varrho$
in the formula(\ref{T28}).
The result could be made more explicit by actually solving equation (\ref{T23})  for $A_0$.
In fact, we shall give a fully explicit formula in the next subsection.

One sees from (\ref{T9}) that for fixed $\xi=\Xi(B)\in \cA_y^\reg$ the set of gauge equivalence classes
\be
\{ [ (A g(v)^{-1} \varrho g(v), B)] \,\vert \, \varrho \in \mathrm{S}\bT^n\}
\label{T29}\ee
is an orbit of the $\beta$-generated torus action.
Thus the above construction implies the \emph{transitivity} of the torus action on
$\hat \beta^{-1}(\xi)$ for all $\xi$ in the image (\ref{T13}).

The next lemma
provides a  characterization of  the stability subgroups for the $\beta$-generated torus action on
$\hat\beta^{-1}(\cA^\reg)$.

\medskip
\noindent
{\bf Lemma 3.} \emph{Consider $\xi$ from the image (\ref{T13}) and
define $\bT^n[v] < \bT^n$ to be the subgroup whose
elements have $v:= v(\xi,y)$ (\ref{T27}) as their eigenvector.
Take an element $[(A,B)] \in P(\mu_0(y))$ that verifies $\Xi(B)=\xi$. Then
$\hat \Psi^\beta_\tau([(A,B)] = [(A,B)]$
holds for precisely those $\tau\in \bT^{n-1}$ for which
\be
\varrho(\tau) =
(g A^{-1} g^{-1}) \zeta (g A g^{-1}) \zeta^{-1}
\quad\hbox{with some}\quad
\zeta\in \bT^n[v].
\label{T30}\ee
Here $(A,B)$ is a representative of $[(A,B)]$, $g$ is any
unitary matrix subject to $g B g^{-1} = \delta(\xi)$ and
$\varrho(\tau)$ refers to (\ref{T10}).
The mapping $\zeta \mapsto \varrho(\tau)$ defines a homomorphism from $\bT^n[v]$
onto the stabilizer subgroup of $[(A,B)]$ with respect to the $\beta$-generated torus action,
 whose kernel
is given by the scalar matrices in $\bT^n$.}

\medskip
\noindent
{\bf Proof.}
Suppose that $[(A,B)]$ is fixed by
$\varrho:= \varrho(\tau)$ (\ref{T10}).
Choosing a representative $(A,B)$, this is equivalent to the existence
of some $h\in G_{\mu_0(y)}$ that satisfies
\be
(A g^{-1} \varrho g, B) = ( h A h^{-1}, h B h^{-1}).
\label{T31}\ee
  Allowing $h$ to be in $\mathrm{U}(n)_{\mu_0(y)}$, the second component says that
\be
h= g^{-1} \zeta g
\label{T32}\ee
for some $\zeta \in \bT^n$.
It is easily seen that $h$ (\ref{T32}) belongs to the little group of $\mu_0(y)$ if and only if
$v(\xi,y)$ is an eigenvector of the diagonal matrix $\zeta$.
We can then solve the equality
\be
A g^{-1} \varrho g = h A h^{-1} = g^{-1} \zeta g A g^{-1} \zeta^{-1} g
\label{T33}\ee
for $\varrho$ as $\varrho = (g A^{-1} g^{-1}) \zeta (g A g^{-1}) \zeta^{-1}$,
which is just the formula (\ref{T30}).

It remains to show that  the right-hand-side of (\ref{T30}) defines
an element in the stabilizer of $[(A,B)]$
for any $\zeta \in \bT^n[v]$.
For this, recall that the moment map constraint is equivalent to
\be
(g A g^{-1}) \delta(\xi) (g A g^{-1})^{-1} = \mu_{v} \delta(\xi),
\label{T34}\ee
where $v$ is the last column
of $g$ and $\mu_v$ is given by (\ref{T20}).
(By a choice of $g$ we may arrange that $v=v(\xi,y)$ (\ref{T27}), but this is inessential:
all vectors whose components have the same absolute values are eigenvectors of the
same diagonal unitary matrices.)
Conjugating this equation by $\zeta$ that has $v$ is its eigenvector, we see that
\be
(g A g^{-1}) \delta(\xi) (g A g^{-1})^{-1} =
(\zeta (g A g^{-1})) \delta(\xi)   (\zeta (g A g^{-1}))^{-1},
\label{T35}\ee
which implies that
\be
\zeta g A g^{-1} = g A g^{-1} \eta(\zeta)
\label{T36}\ee
for some $\eta(\zeta)\in \bT^n$.
Therefore  $\varrho :=\eta(\zeta) \zeta^{-1}$ is also diagonal, and it
 belongs to the stabilizer of $[(A,B)]$ since (\ref{T36}) implies
(\ref{T31}) with $h:= g^{-1} \zeta g \in \mathrm{U}(n)_{\mu_0(y)}$.

It is readily verified that the map $\zeta \mapsto \varrho(\tau)$ (\ref{T30}) is a homomorphism,
which does not depend on the choices (of $(A,B)$ and $g$) that were made
in its construction.
To finish the proof, suppose that $\zeta$ is in the kernel of this homomorphism.
This means that
\be
\zeta A^g \zeta^{-1} = A^g
\quad\hbox{for}\quad A^g:= g A g^{-1},
\label{T37}\ee
and since $ g B g^{-1}= \delta(\xi)$ we conclude that
$g^{-1} \zeta g$
fixes $(A,B)$ by the componentwise conjugation action.
Since we know \cite{FK} that $\mathrm{U}(n)_{\mu_0(y)}/\mathrm{U}(1)$ acts freely
on $\mu^{-1}(\mu_0(y))$, we obtain that
$g^{-1} \zeta g$ must belong to the scalar matrices
$\mathrm{U}(1) < \mathrm{U}(n)$, and hence $\zeta$
has the same property.
\qed

Those vectors $v(\xi,y)$ that have only non-vanishing components are eigenvectors
of the scalar elements of $\bT^n$ only, and therefore the $\beta$-generated torus action is \emph{free} on the
corresponding fibres $\hat \beta^{-1}(\xi)$ (for example on
$\hat \beta^{-1}(\xi^*)$ with $\xi^*$ in (\ref{T25})).
In particular, this shows that the torus action is effective.
 On the other hand, using the fact that the common eigenvectors
 of $\bT^n$ are those vectors
that have a single non-zero component, Lemma 3 implies the following useful statement.

\medskip
\noindent
{\bf Corollary 4.}
\emph{
The fixed points of the $\beta$-generated torus action are in one-to-one correspondence
with the set of $\xi \in \cA^\reg$ for which the equations
\be
z_\ell(\xi,y) = \delta_{k,\ell},
\qquad
\ell = 1,\ldots, n
\label{T38}\ee
hold for some arbitrarily fixed $k\in \{ 1,2,\ldots, n\}$. }

\medskip
\noindent
{\bf Remark  5.}
The centre $\bZ_n:= \bZ/n\bZ$ of $\mathrm{SU}(n)$ acts on $\cA$ as given by the following
action of the generator $\sigma$:
\be
\sigma(\xi)_k:= \xi_{k+1}.
\label{cyclic}\ee
One can check from (\ref{T15}) that
\be
z_\ell(\sigma(\xi), y) = z_{\ell +1}(\xi,y),
\qquad
\forall \xi\in \cA^\reg,
\label{sigma}\ee with the
convention $z_{\ell +n}:= z_\ell$.
It follows that
the image $\cA_y^\reg$ (\ref{T13}) as well as the set of fixed points
of the $\beta$-generated torus action
are
invariant under this action of $\bZ_n$. Moreover,  Corollary 4 implies that
the $\bZ_n$-orbit of any chosen fixed point of the torus action
 consists of $n$ different fixed points.
 By noting that the $\bZ_n$-action engendered by (\ref{cyclic}) is inherited from the action
 of the centre of $\mathrm{SU}(n)$ on $\mathrm{SU}(n)$ by left-multiplications,
 it is readily seen that
 the full image
 \be
 \cA_y:= \hat\beta(P(\mu_0(y))
 \label{Ay}\ee
  is also mapped to itself by $\sigma$.

\subsection{RS system on dense open submanifold of $P(\mu_0(y))$}

We show below that the reduction leads to an integrable system whose
``principal Hamiltonian'' takes the  RS form (\ref{I14}) on a dense open submanifold
of the reduced phase space.
For our characterization of this system,  it will be useful
to decompose $\cA_y$ (\ref{Ay}) into the union of 3 disjoint subsets:
\be
\cA_y = \cA_y^\reg \cup \cA_y^\sing = \cA_y^+ \cup \cA_y^= \cup \cA_y^\sing,
\label{Adec}\ee
where $\cA_{ y}^\sing:= \cA_y \cap \partial \cA$ and
\be
\cA_{y }^+ := \{ \xi \in    \cA^\reg\,\vert z_\ell(\xi,y) >0,
\quad \forall \ell=1,\ldots,n\,\},
\label{T44}\ee
\be
\cA_{y}^= := \{ \xi \in    \cA^\reg\,\vert z_\ell(\xi,y) \geq 0,
\quad \forall \ell=1,\ldots,n, \quad \prod_{\ell=1}^n z_\ell(\xi,y)=0\}.
\label{T44=}\ee
Their significance is that the $\beta$-generated torus action is free on
$\hat \beta^{-1}(\cA_y^+)$, has non-trivial isotropy  groups on $\hat\beta^{-1}(\cA_y^=)$,
and is not defined at all on $\hat\beta^{-1}(\cA_y^\sing)$ (which is empty
if (\ref{T2}) holds).
It turns out that these sets depend only on the absolute value of $y\in (-\pi/2, \pi/2)$,
and each of them is mapped to itself by the cyclic permutation $\sigma$ (\ref{sigma})
and the ``partial reflection'' $\nu$ that maps $\xi$ to $\nu(x)$ according to
\be
\nu(\xi)_k = \xi_{n-k}
\quad
\forall k =1,\ldots, n-1
\quad\hbox{and}\quad
\nu(\xi)_n = \xi_n.
\label{nu}\ee

In order to derive the above mentioned properties,
we begin by pointing out that the equalities
\be
\alpha(\mu^{-1}(\mu_0)) = \beta(\mu^{-1}(\mu_0)) = \alpha(\mu^{-1}(\mu_0^{-1})) = \beta(\mu^{-1}(\mu_0^{-1}))
\label{T39}\ee
are valid for any moment map value $\mu_0\in G$.
To see this, first remark \cite{FK} that
\be
(A,B) \in \mu^{-1}(\mu_0) \Longleftrightarrow S(A,B) := (B^{-1}, B A B^{-1}) \in \mu^{-1}(\mu_0).
\label{T40}\ee
On $\mu^{-1}(\mu_0)$ we thus have
\be
\alpha = \beta \circ S,
\label{T41}\ee
and since $S$ is a diffeomorphism of $\mu^{-1}(\mu_0)$ this entails that the $\alpha$-image of
$\mu^{-1}(\mu_0)$ is the same as its $\beta$-image.
Second, by inverting the group commutator, notice that
\be
(A,B) \in \mu^{-1}(\mu_0) \Longleftrightarrow (B,A) \in \mu^{-1}(\mu_0^{-1}),
\label{T42}\ee
which implies  the second equality in (\ref{T39}).

Since $\mu_0(y)^{-1} = \mu_0(-y)$, we conclude from the above that
\be
\beta(\mu^{-1}(\mu_0(y))) = \beta(\mu^{-1}(\mu_0(-y))).
\label{T43}\ee
We also observe from (\ref{T14}) that if $\xi$ is such an element of
$\cA^\reg$ for which $z_\ell(\xi,y)$ is non-zero
for all $\ell=1,\ldots, n$, then $\xi$ verifies the same property for $-y$.
Taking advantage of the  identity
$\Xi_k(\delta(\xi)^{-1}) = \nu(\xi)_k$ and componentwise complex conjugation of the pair
$(A,B) \in \mu^{-1}(\mu_0(y))$,
it follows that $\cA_y=\cA_{-y}$ (\ref{Ay}) is stable under the involution $\xi\mapsto \nu(\xi)$.

We now focus on the subset of $P(\mu_0(y))$
given by the  inverse image $\hat \beta^{-1}(\cA_y^+)$.
Note that $\xi^*$ (\ref{T25}) always belongs to $\cA_{y}^+$, which
is therefore a non-empty open subset of $\cA^\reg$.
Since $\hat \beta$ is continuous, $\hat \beta^{-1}(\cA_y^+) \subset P(\mu_0(y))$ is
a non-empty open submanifold.

Define the smooth matrix function $\cL_y^\loc$ on $\cA_{y}^+ \times \bT^{n-1}$
by the formula
\be
\cL_y^\loc(\xi,\tau)_{j \ell} :=
\frac{\sin (ny)}{\sin (y)}
\frac{e^{\ri y} - e^{-\ri y}}{e^{\ri y} \delta_j(\xi) \delta_\ell(\xi)^{-1} - e^{-\ri y}}
v_j(\xi,y) v_\ell(\xi,-y) \varrho(\tau)_\ell.
\label{T45}\ee
Further, taking any vector $v\in \bR^n$ that has unit norm and component $v_n\neq -1$,
introduce the  unitary matrix
$g(v)\in \mathrm{U}(n)$  by
\bea
&&g(v)_{jn} := - g(v)_{nj}:= v_j,
\quad
\forall j=1,\ldots, n-1,
\quad
g(v)_{nn} := v_n,\nonumber\\
&& g(v)_{jl} := \delta_{jl}- \frac{v_j v_l}{ 1 + v_n},
\quad
\forall j,l=1,\ldots, n-1.
\label{T46}\eea
Then set
\be
g_y(\xi):= g(v(\xi,y)),
\qquad
\forall \xi\in \cA_{y}^+,
\label{T47}\ee
where $v(\xi,y)$ denotes the positive vector
$v_\ell(\xi,y)= \sqrt{z_\ell(\xi,y)}$.

We are ready to present the main result of this section, which
generalizes Theorem 4 of \cite{FK}.

\medskip \noindent
{\bf Theorem 6.} \emph{For any $y\in (-\pi/2, \pi/2)$ subject to (\ref{T1}),  the set of elements
\be
\left\{\left( g_y(\xi)^{-1} \cL_y^\loc(\xi,\tau) g_y(\xi), g_y(\xi)^{-1}
\delta(\xi)g_y(\xi) \right)\,\Big\vert \,
(\xi,\tau)\in \cA^+_{y} \times \bT^{n-1}\,\right\}
\subset G \times G
\label{T48}\ee
defines a cross-section of the orbits of $G_{\mu_0(y)}$
in the open submanifold
$\beta^{-1}(\cA^+_{y}) \cap \mu^{-1}(\mu_0(y))$
of the constraint surface.
The one-to-one parametrization of this cross-section by $(\xi,\tau) \in
\cA^+_{y} \times \bT^{n-1}$ induces Darboux coordinates
on the corresponding open submanifold of the reduced phase space,
\be
 \hat \beta^{-1}( \cA^+_{y }) \subset P(\mu_0(y)) = \mu^{-1}(\mu_0(y))/ G_{\mu_0(y)},
\label{T49}\ee
  since on this submanifold the symplectic form that descends from  $\omega^\lambda$ in (\ref{I4})
can be written as
\be
\omega_\red^\loc
=\ri \lambda \sum_{k=1}^{n-1} \mathrm{d} \xi_k
\wedge \mathrm{d} \tau_k \tau_k^{-1} = \lambda \sum_{k=1}^{n-1} \mathrm{d} \theta_k \wedge \mathrm{d} \xi_k
\quad\hbox{with}\quad \tau_k = e^{\ri \theta_k}.
\label{T50}\ee
The submanifold $\hat \beta^{-1}( \cA^+_{y})$ is a dense subset of the full reduced phase space.
On this submanifold the  Poisson commuting reduced Hamiltonians
descending from the smooth class functions of $A$ in $(A,B)\in G\times G$ are given by the class functions
of the $\mathrm{SU}(n)$-valued ``local Lax matrix'' $\cL_y^\loc(\xi,\tau)$.
In particular, using
$s:= \operatorname{sign}\!\left( \frac{\sin(y)}{\sin(ny)}\right)$ and $\theta_0=\theta_n:=0$,
the reduction of the function $\Re\!\left(\tr( A)\right)$
yields the generalized RS Hamiltonian
 \be
H_{ y }^\loc(\xi,\theta) := \Re\!\left(\tr
\left(\cL_y^\loc(\xi,\tau)\right)\right) =
 s\sum_{j=1}^n \cos(\theta_j - \theta_{j-1}) \prod_{k=j+1}^{j+n-1}
\left\vert 1 - \frac{\sin^2 y}{ \sin^2(\sum_{m=j}^{k-1} \xi_m)} \right\vert^\frac{1}{2}.
\label{T51}\ee
}

\medskip
The first statement of the theorem requires proving that the set (\ref{T48}) lies
in the ``constraint surface'' $\mu^{-1}(\mu_0(y))$ and its intersection with any orbit of
$G_{\mu_0(y)}$ consists of at most one point.
The second statement requires calculation of the pull-back of the
quasi-Hamiltonian 2-form (\ref{I4}) on the set (\ref{T48}).
The proof of both parts follows word-by-word the proof of the corresponding
statements of Theorem 4 of \cite{FK}, and hence is omitted.

The proof of the denseness statement is trivial if
(\ref{T2}) holds, i.e., if  $\cA_y^\sing = \emptyset$. In such cases  $P(\mu_0(y))$ is a Hamiltonian toric manifold
under the $\beta$-generated torus action,
and $\hat \beta^{-1}( \cA^+_{y})$ gives the corresponding submanifold of principal orbit type,
which is known to be dense and open.
Regarding the cases when $\cA_y^\sing \neq \emptyset$, the  denseness
is proved in Appendix B.
Finally, the formula (\ref{T51}) follows by straightforward calculation.

We finish this section with a few comments.
First of all, we recall that
in the case of the regime (\ref{I2})
the Hamiltonian (\ref{T51}) is just the
 standard RS Hamiltonian of $\IIIb$ type \cite{RIMS95}.
The principal message of the theorem is that
\emph{the local RS Hamiltonian defined by (\ref{T51}) on the domain
$\cA^+_{y} \times \bT^{n-1}$  extends uniquely
to a globally smooth Hamiltonian on the compact reduced phase space
$P(\mu_0(y))$ for  any parameter $y\in (-\pi/2, \pi/2)$ subject to (\ref{T1}). }

The  domain $\cA_y$ is in general
different from  the Weyl alcove with thick walls (\ref{I10}).
We shall investigate the dependence of this domain
on $ y $ in the following section. Here
it is worth noting that the continuity of $\hat \beta\colon P(\mu_0(y)) \to \cA$
and the denseness statement in Theorem 6 imply that $\cA_y^+$ is always a dense subset of $\cA_y$.

By the duality between the functions $\hat\alpha^\lambda$ and $\hat \beta^\lambda$,
which arises from the relation (\ref{T41}), the components of $\hat \alpha^\lambda$
generate a free Hamiltonian torus action on the dense open submanifold
$\hat\alpha^{-1}(\cA_y^+)\subset P(\mu_0(y))$.  This shows the Liouville integrability of the
commuting set of globally smooth Hamiltonians that descend from the smooth class function
of the matrix $A$ in $(A,B)\in D$.

\section{Classification of the coupling parameter}
\setcounter{equation}{0}

We have seen that our reduction always yields a Liouville integrable system whose
 leading Hamiltonian  has the RS form of $\IIIb$ type (\ref{T51}) on a dense open submanifold
 of the compact reduced phase space $P(\mu_0(y))$.
In principle,
two different types of cases can occur:
\begin{itemize}

\item{Type (i): the constraint surface satisfies $\mu^{-1}(\mu_0(y))\subset G_\reg \times G_\reg$.}

\item{Type (ii): the relation $\mu^{-1}(\mu_0(y)) \subset G_\reg \times G_\reg$ does not hold.}

\end{itemize}
In the type (i) cases the reduced phase space inherits globally smooth action and position variables from the double.
In the type (ii) cases
neither the action variables nor the  position variables
extend to globally smooth (differentiable) functions on the full reduced
phase space $P(\mu_0(y))$.
This follows from the fact that the components of the spectral function $\Xi$ (\ref{T3}),
whereby $\alpha$ and $\beta$ (\ref{T4}) descend to
action variables  and position variables,
develop singularities at  the non-regular elements of $G$,
and those singularities cannot disappear by the reduction.
It is also worth noting  that
at non-regular elements the dimension of the span
of the  differentials
of the smooth class functions of $G=\mathrm{SU}(n)$
 is always smaller than $(n-1)$.
These group-theoretic results are elucidated in Appendix A.

In this section we show that both type (ii) and new type (i) cases exist, and give
the precise classification of the coupling parameter $y$ according to this dichotomy.
Moreover, we shall prove that in the type (i) cases the full reduced phase space is always
symplectomorphic to $\cp$ with a multiple of the Fubini-Study form.
The final results are given by Theorems 12 and 13 below.

Using that $y$ matters only modulo $\pi$,
we here parametrize $\mu_0(y)$  by $y$ taken from the range
\be
0<y <\pi.
\label{F1}\ee
It is proved in Appendix B that the $\beta$-image $\cA_y$ of the constraint surface
is the closure of $\cA_y^+$ defined in (\ref{T44}).
Now the domain $\cA_y^+$ can be characterized as follows.

\medskip
\noindent
{\bf Theorem 7.} \emph{Take any $y$  subject to (\ref{T1}), (\ref{F1}) and let $k\in \{0,\ldots, n-1\}$
be the integer
verifying
\be
k \pi/n  < y < (k+1) \pi/n.
\label{F2}\ee
Then $\cA_y^+$ consists of those elements $\xi\in \cA^\reg$ whose components
satisfy the following condition for each $\ell=1,\ldots, n$:
\be
\xi_\ell > y
\quad
\hbox{if}\quad k=0,
\label{F3}\ee
 \be
\xi_{\ell}+\cdots+\xi_{\ell + k -1} < y \quad\hbox{and}\quad
\xi_{\ell}+\cdots+\xi_{\ell + k} >  y
\quad \hbox{if}\quad k=1,\ldots, n-2,
\label{F4}
\ee
\be
\xi_{\ell}+\cdots+\xi_{\ell + n -2} < y
\quad \hbox{if}\quad k=n-1.
\label{F5}\ee
}
\medskip
\noindent
{\bf Proof.}
Recall from (\ref{T44}) that $\xi \in \cA_y^+$ if and only if $z_\ell(\xi,y)>0$ for each
$\ell=1,\ldots,n$.
By inspecting the formula (\ref{T15}) one sees that $z_\ell(\xi,y)>0$ holds if and
only if $\xi$ satisfies the inequalities
\be
\xi_{\ell}+\cdots+\xi_{\ell + \kappa(\ell) -1} < y
\label{F6}\ee
and
\be
\xi_{\ell}+\cdots+\xi_{\ell + \kappa(\ell)} >  y
\label{F7}\ee
for some
\be
\kappa(\ell)\in \{ 0,1,\ldots, n-1\} \quad\hbox{subject to}\quad (-1)^{\kappa(\ell)} = (-1)^k.
\label{F8}\ee
The above inequalities say that the number of  $\xi$-dependent
negative factors in the product that gives $z_\ell(\xi,y)$ (\ref{T15}) is $\kappa(\ell)$,
while rest of the  factors is positive.
We utilized that,
 on account of (\ref{F2}),
the sign of the ``pre-factor''  $\sin(y)/\sin(ny)$ in (\ref{T15}) is the same as the sign of  $(-1)^k$.

The sums in (\ref{F6}) and in (\ref{F7}) contain  $\kappa(\ell)$ and
$(\kappa(\ell)+1)$ terms, respectively. If $\kappa(\ell)=0$, then equation
(\ref{F6}) is absent (automatic if  the value of the empty sum is taken to be zero),
 and if $\kappa(\ell)=(n-1)$, then equation (\ref{F7}) holds automatically.
In principle, $\kappa(\ell)$ could be a non-constant function of $\ell$ and it
 could also vary as $\xi$ varies.

 We now demonstrate that the  inequalities (\ref{F6}) and (\ref{F7})  together with (\ref{F8}) enforce that
 \be
 \kappa(\ell) = k.
 \label{F9}\ee
 To this end, we first show that the relations
\be
\kappa(\ell) \leq \kappa(\ell+1)
\quad\hbox{for}\quad
\ell=1,\ldots, n-1
\label{F10}\ee
and
\be
\kappa(n) \leq \kappa(1)
\label{F11}\ee
must hold, which  entail that $\kappa$ is an $\ell$-independent constant.
We remark that the formula of the function $z_\ell$  can be extended by
periodicity,  $z_{\ell +n}= z_\ell$, and then one must also have
$\kappa(\ell+n) = \kappa(\ell)$.  With this convention, (\ref{F11}) is
just the $\ell=n$ instance of (\ref{F10}).

To derive (\ref{F10}), fix some $1\leq \ell\leq (n-1)$ and note first that
if $\kappa(\ell) \in \{ 0,1\}$, then (because of (\ref{F8})), there is nothing to prove.
Suppose then that $\kappa(\ell) \geq 2$ and suppose also that (\ref{F10}) does not hold
at this $\ell$.
 Since the parity of $\kappa(\ell)$ is independent of $\ell$, this means that
\be
\kappa(\ell +1) \leq \kappa(\ell)-2,
\label{F12}\ee
which is equivalent to
\be
\ell + 1 + \kappa(\ell +1) \leq \ell + \kappa(\ell)-1.
\label{F13}\ee
But then we would obtain that
\be
\xi_{\ell+1} + \cdots + \xi_{(\ell + 1) + \kappa(\ell +1)} \leq
\xi_{\ell+1} + \cdots + \xi_{\ell + \kappa(\ell)-1} \leq
\xi_{\ell} + \xi_{\ell+1} + \cdots + \xi_{\ell + \kappa(\ell)-1}.
\label{F14}\ee
This is a contradiction since the first sum in (\ref{F14}) is larger than $y$ by (\ref{F7})
applied to $(\ell+1)$, while the last sum is smaller than $y$ by (\ref{F6}) applied to $\ell$.

Invoking the periodicity, $z_\ell=z_{\ell +n}$,  or by direct inspection of (\ref{F6}) and (\ref{F7})
for $\ell =n$ and $\ell =1$, we obtain equation (\ref{F11}) as well.

Let $\kappa_0$ denote  the value of the constant $\kappa(\ell)$.
By taking the sum of the respective inequalities in (\ref{F6}) and (\ref{F7})
 for $\ell=1,\ldots, n$, we see that
\be
\kappa_0 \pi < ny
\quad\hbox{and}\quad
(\kappa_0 + 1) \pi > n y.
\label{F15}\ee
Comparison with (\ref{F2}) shows that $\kappa_0 = k$, whence the proof is complete.
\qed

\medskip

The type (i) cases are precisely those for which $\cA_y$ does not intersect the
boundary $\partial \cA$ of the alcove $\cA$ (\ref{I7}).
 The subsequent analysis will lead to a complete description
 of the $y$-values when this holds.
To start, introduce the affine space $E$ by
\be
E := \{ \xi \in \bR^n \mid \xi_1 + \cdots + \xi_n = \pi \}.
\label{F16}\ee
Then, for any integer $1\leq p \leq (n-1)$
and $0< y < \pi$ not equal to $p\pi/n$, define the closed convex polyhedron as the subset of $E$
 given by requiring the following:
\begin{itemize}
 \item The bounding hyperplanes of $\cB(p,y)$ are defined by the $n$
 cyclic permutations of the equation
 \be
 \xi_1 + \cdots + \xi_p = y.   \label{eq:face-of-B}
 \ee
 \item The polyhedron $\cB(p,y)$ contains the point $\xi^*$ (\ref{T25}).
\end{itemize}
We additionally define $\cB(0,y) = \cB(n,y) = E$ and also let $\cB(p,y)^\circ$ denote the interior
of $\cB(p,y)$.
We remark that $\cB(p,y)$ is not necessarily bounded.

With the above definitions, we have
\be
\cA_y =  \cB(k,y) \cap \cB(k+1,y)
\quad\hbox{if}\quad
k \pi/n < y < (k+1)\pi/n, \quad k=0,\ldots, n-1.
 \label{eq:A-cap-B-cap-B}
\ee
Indeed, if (\ref{F2}) holds then  $\cB(k,y)$ and $\cB(k+1,y)$ are respectively given by imposing
\be
\xi_{\ell}+\cdots+\xi_{\ell + k -1} \leq y
\quad\hbox{and}\quad
\xi_{\ell}+\cdots+\xi_{\ell + k} \geq  y,\quad
\forall \ell=1,\ldots, n,
\label{F19}\ee
on $\xi \in E$.
The differences
of these equations imply that $\xi_{\ell+k}\geq 0$ for all $\ell$,
i.e., the intersection on the right hand side of
(\ref{eq:A-cap-B-cap-B}) lies in $\cA$.
Thus (\ref{eq:A-cap-B-cap-B}) follows from Theorem 7 and from the fact that
$\cA_y$ is the closure of $\cA_y^+$.
One should note that $\cA_y$ is of interest only under the additional regularity condition (\ref{T1}) on $y$,
but below it will be
convenient to formulate various statements for slightly more general values of $y$.

Let us consider the finite ring $\bZ/n\bZ$.
Addition and multiplication in $\bZ/n\bZ$ are inherited from $\bZ$,
and we choose to represent the equivalence classes by
$\{1,2,\ldots, n\}$.
It is well-known that
if $n$ and $1\leq p\leq (n-1)$ are relatively prime, $\gcd(n,p)=1$, then multiplication by $p$
gives a permutation
of the elements of this ring.
In particular, there exits a unique integer $1\leq q \leq (n-1)$  such that
$pq = 1 \mod n$. This will be crucial in proving the following lemma, which exhibits cases when
$\cB(p,y)$ is bounded.

\medskip
\noindent
{\bf Lemma 8.} \emph{If the integers $1\leq p \leq (n-1)$ and $n$ are relatively prime
and  $0<y<\pi$ satisfies
 $y \neq p \pi/n$, then $\cB(p,y)$ is an $(n-1)$-dimensional simplex.
Writing $q$ for the integer  $1 \leq q \leq n-1$ such that $pq = 1 \mod n$, and defining
\be
\tilde{y} := y - \frac{p \pi }{n},\quad
a := \frac{\pi}{n} + q\tilde{y},\quad
b := \frac{\pi}{n} - (n-q)\tilde{y}, \label{eq:a-b-for-vertex-of-B}
\ee
the $n$ vertices of $\cB(p,y)$ are the cyclic permutations of the point $x\in E$ given by
\bea
x_i = a &\quad& \mbox{for } i = j p \mbox{ with } j = 1,\ldots,n-q, \nonumber \\
x_i = b &\quad& \mbox{otherwise,}  \label{eq:vertex-of-B}
\eea
where the index $i$ is read modulo $n$.}

\medskip
\noindent
{\bf Proof.}
If  the polyhedron $\cB(p,y)$ is bounded, then it must be a simplex,
since it is bounded by $n$ hyperplanes in the $(n-1)$-dimensional space $E$ and
contains a neighborhood of the point $\xi^*$.
One knows from the Minkowski-Weyl theorem \cite{Webster} that $\cB(p,y)$ is not bounded if and only if it
contains a half-line, i.e., a set of elements of the form
\be
 \xi(\lambda) = c + \lambda d,
 \quad
 \forall \lambda \geq 0,\quad   E \ni d\neq 0.
 \label{F22}\ee
 We next show that such a half-line does not exist.

 Let $e_i$ ($i=1,\ldots, n)$ be the standard basis of $\bR^n$ and apply
the  convention $e_j = e_{j\pm n}$ for all $j\in \bZ$.
Define $\epsilon:= e_1 +\cdots + e_n$ and
\be
V_i(p):= e_i + e_{i+1} + \cdots + e_{i+p-1},
\quad\forall i\in \bZ.
\label{F23}\ee
Supposing for definiteness that $y> p \pi/n$,
$\cB(p,y)$ consists of the elements $x\in \bR^n$ for which
\be
 \epsilon\cdot x =\pi \quad\hbox{and}\quad
V_i(p)\cdot x \leq y, \quad \forall i.
\label{F24}\ee
Therefore, the direction vector $d$ of a half-line contained in $\cB(p,y)$ must satisfy
\be
 d \cdot \epsilon  =0
 \quad
 \hbox{and}\quad
 d\cdot  V_i(p) \leq 0, \quad \forall i.
\label{F25} \ee
Since $V_1(p) +\cdots +V_n(p)= p \epsilon$, these conditions imply
\be
d\cdot V_i(p) =0,
\quad
\forall i.
\label{F26}\ee
Let us expand the vector $d$ as
 \be
 d = d_1 e_1 +\dots +d_n e_n
 \label{F27}\ee
 and set $d_j := d _{ j \pm n}$ for all $j\in \bZ$.
 By writing
 \be
 p q = r n + 1
 \quad\hbox{with some} \quad r\geq 0,
 \label{F28}\ee
 one has the identity
 \be
 V_i(p) + V_{i + p}(p) + \cdots + V_{i + (q-1)p}(p) = r \epsilon + e_i,
 \quad
 \forall i.
 \label{F29}\ee
 Taking the scalar product of this identity with $d$, using that $d\cdot \epsilon =0$, leads to
 \be
 d_i = e_i \cdot d = 0,
 \quad
 \forall i.
 \label{F30}\ee
 Hence the polyhedron $\cB(p,y)$ contains no half-line.
 A similar argument works also if $y < p \pi/n$.

 Now that we know that $\cB(p,y)$ is a simplex,  we need to calculate its vertices.
 Since the $n$ vertices are clearly the cyclic permutations of a single one, it is
 enough to find the vertex $x$ that solves the first $n-1$ cyclic permutations of equation (\ref{eq:face-of-B}).
 Taking subsequent differences of these $n-1$ equations gives the relations
\be
x_i = x_{i+p} \quad \hbox{for }\quad i=1,\ldots,n-2,
 \label{eq:equivalence-relation-on-coeffs}
\ee
where the indices are understood modulo $n$.
The assumption $\gcd(n,p)=1$ implies that for each $i=1,\ldots, n-2$ there exists
a unique $m_i \in \{1,\ldots, n-1\} \setminus \{ (n-q)\}$ such that $i = m_i p \mod n$,
where we used that $n-1= (n-q) p \mod n$.
It follows immediately that the relations (\ref{eq:equivalence-relation-on-coeffs})
can be recast in the form
\bea
&&x_p = x_{2p} = \cdots = x_{(n-q)p} := a, \nonumber\\
&&x_{(n-q+1)p} = x_{(n-q+2)p} = \cdots = x_{np} := b,
\label{F32}\eea
with some constants $a$ and $b$.

We are left with the task of calculating $a$ and $b$. We have two linear equations for this task.
First of all, the condition $x \in E$ is equivalent to
\be
q b + (n-q)a = \pi.
\label{F33}\ee
To obtain the second  equation, we sum \emph{all} cyclic permutations of (\ref{eq:face-of-B}) for $x$.
On the one hand,
this sum contains each coefficient $p$ times, so (by $x\in E$) it must be equal to $\pi p$. On the other hand,
notice that for the $n$-th cyclic permutation that was omitted we have
\bea
x_n + x_1 + x_2 + \cdots +x_{p-1} &=& x_n - x_p + (x_1 + \cdots + x_p) \nonumber \\
                                  &=&  b  -  a  + (x_1 + \cdots + x_p) \nonumber \\
                                  &=&  b  -  a  + y.   \label{eq:last-cyclic-permutation}
\label{F34}\eea
Therefore, summing all cyclic permutations gives
\be
\pi p = n y + b - a.
\label{F35}\ee
Equations (\ref{F33}) and (\ref{F35}) for $a$ and $b$ are solved uniquely by the formula
(\ref{eq:a-b-for-vertex-of-B}).
 \qed

\medskip

One sees from Lemma 8 that as $y$ approaches $p\pi/n$ the simplex $\cB(p,y)$
contracts onto the point $\xi^*$.
Then as $y$ moves away from $p\pi/ n$  the simplex grows and at some value of $y$  its
vertices reach $\partial\cA$.
The range of $y$ for which it stays inside the interior $\cA^\reg$ of $\cA$ is
described as follows.

\medskip
\noindent
{\bf Corollary 9.} \emph{For $\gcd(n,p) = 1$, the simplex $\cB(p,y)$ is contained in $\cA^\reg$ if and only if
$y\neq \frac{ p\pi }{n}$ belongs to the open interval
$\left(\frac{p \pi }{n}  -\frac{\pi}{nq}, \frac{p\pi}{n} +\frac{\pi}{(n-q)n}\right)$,
where $q$ is defined as in Lemma 8.}

\medskip
\noindent
{\bf Proof.} The simplex $\cB(p,y)$ is contained in $\cA^\reg$ if and only if its vertices are
contained in  $\cA^\reg$, which means that both $a$ and $b$ in
(\ref{eq:a-b-for-vertex-of-B}) are positive.
If $y > p \pi/n$, then $a>0$ and the positivity of $b$ is equivalent to $y < p \pi/n + \pi/(n(n-q))$.
If $y< p \pi/n$, then $b>0$ and the positivity of $a$ is equivalent to $y > p \pi/n - \pi/(nq)$.
\qed

\medskip
\noindent
{\bf Lemma 10.} \emph{Suppose $\gcd(n,p) = 1$ and take $y$ from the interval
given in Corollary 9
such that it is not an integer multiple of $\pi/n$. In this case the simplex
$\cB(p,y)\subset \cA^\reg$ verifies the following property.
\begin{itemize}
 \item
 If
 $\frac{p\pi}{n} <y <\frac{p\pi}{n} +\frac{\pi}{(n-q)n}$, then $\cB(p,y) \subset  \cB(p+1,y)^\circ$.
 \item
 If
 $\frac{p\pi}{n} -\frac{\pi}{nq} < y < \frac{p\pi}{n}$,  then $\cB(p,y)\subset \cB(p-1,y)^\circ$.
\end{itemize}
}

\medskip
\noindent
{\bf Proof.}
Let us pick a vertex $x$ of the simplex $\cB(p,y)\subset \cA^\reg$ and recall that it satisfies all
 but one of the $n$  cyclic permutations of the equation
\be
x_1 + \cdots + x_p =y.
\ee
In particular, it satisfies at least one of the following two equations
\be
x_\ell + \cdots + x_{\ell + p -1} =y
\quad\hbox{or}\quad
x_{\ell + 1} + \cdots +  x_{\ell+ p} = y
\label{P12}\ee
for each $\ell =1,\ldots, n$.
Suppose now that $\frac{p\pi}{n} <y <\frac{p\pi}{n} +\frac{\pi}{(n-q)n}$, which entails
that the polyhedron $\cB(p+1,y)^\circ$ is given by the inequalities
\be
\xi_\ell + \cdots + \xi_{\ell + p} > y.
\ee
The fact that all components of $x$ are positive implies by (\ref{P12}) that the vertex
$x$ of $\cB(p,y)$ lies in $\cB(p+1,y)^\circ$.
The case $\frac{p\pi}{n} -\frac{\pi}{nq} < y < \frac{p\pi}{n}$ is settled quite similarly
by using that in this case the defining inequalities of $\cB(p-1,y)^\circ$ are
$\xi_\ell +\cdots + \xi_{\ell + p-2} <y$ if $p>1$ and $\cB(0,y)^\circ=E$.
\qed

\medskip
\noindent
{\bf Proposition 11.} \emph{Let $n\geq 2$ be given and pick $1 \le p \le n-1$
such that $\gcd(n,p)=1$. Define $q$ as
in Lemma 8  and consider
$y\in \left(\frac{\pi p}{n}  -\frac{\pi}{nq},
\frac{p\pi}{n} +\frac{\pi}{(n-q)n}\right)$ subject to (\ref{T1}).
Then  $\cA_y = \cB(p,y)$.}

\medskip
\noindent
{\bf Proof.} This is a direct consequence of the relation (\ref{eq:A-cap-B-cap-B}),
whereby  $\cA_y = \cB(p-1,y) \cap \cB(p,y)$ if  $ p \pi/n  -\pi/nq< y < p\pi/n$ and
$\cA_y = \cB(p,y) \cap \cB(p+1,y)$ if  $ p \pi/n  < y < p\pi/n  +\pi/(n(n-q))$, and
the statement of Lemma 10.
\qed

Now we are ready to formulate the main results of the present paper.

\medskip
\noindent
{\bf Theorem 12.}  \emph{Consider the reduction of the double $\SU(n)\times \SU(n)$
defined by the moment map constraint $\mu(A,B)= \mu_0(y)$ with
$0<y < \pi$ subject to (\ref{T1}).
 Suppose that $y$
belongs to an open interval of the form
\be
\left(\frac{p\pi}{n} - \frac{\pi}{nq}, \frac{p\pi}{n} + \frac{\pi}{(n-q)n}\right),
\label{cpcase}\ee
where $\gcd(n,p)=1$ and $pq = 1 \mod n$ with integers $1\leq p, q \leq (n-1)$.
Then the $\beta$-image $\cA_y$ of the constraint surface $\mu^{-1}(\mu_0(y))$
is contained in $\cA^\reg$.
In these cases
the reduced phase space $P(\mu_0(y))$ is symplectomorphic to $\cp$  with a multiple of
the Fubini-Study symplectic structure. }

\medskip
\noindent
{\bf Proof.}
Proposition 11 and the preceding lemmas ensure that if $y$ satisfies (\ref{T1}) and (\ref{cpcase}), then
the $\beta$-image of the constraint surface is provided by the simplex $\cB(p,y)$, which is
contained in $\cA^\reg$.
This implies that the reduced phase space is a Hamiltonian toric manifold with respect
to the toric moment map $\hat \beta^\lambda$ having the image $\lambda \cB(p,y)$,
where the constant $\lambda$ gives the scale of the quasi-Hamiltonian 2-form (\ref{I4}).
Up to symplectomorphisms, the only toric manifold whose ``Delzant polytope'' is
an $(n-1)$-dimensional simplex is $\cp$ equipped with a multiple of the Fubini-Study symplectic
form \cite{De,Can}.
\qed

\medskip
\noindent
{\bf Theorem 13.} \emph{The values of $y$ given in Theorem 12 exhausts all type (i) cases.  In other words,
if $0<y < \pi$ subject to (\ref{T1}) does not belong to an open interval of the form (\ref{cpcase}),
then  $\cA_y$ intersects the boundary $\partial \cA$ of $\cA$.
}

\medskip
The proof will follow from a few simple lemmas.
First of all, for any $y$ as in equation (\ref{F2}) we let $\cC_y$ denote the set of those $\xi\in \cA$ that
satisfy the inequalities
\be
\xi_{\ell}+\cdots+\xi_{\ell + k -1} \leq y \quad\hbox{and}\quad
\xi_{\ell}+\cdots+\xi_{\ell + k} \geq  y
\ee
for each $\ell =1,\ldots, n$ (where the first inequality
is automatic if $k=0$).
This means that $\cC_y = \cA_y$ if $y$ also satisfies (\ref{T1}).

\medskip
\noindent
{\bf Lemma 14.} \emph{Suppose that
$k \pi/n < y_1 < y_2 < (k+1)\pi/n$ and both $\cC_{y_1} \cap \partial\cA$ and $\cC_{y_2} \cap\partial  \cA$
are non-empty.
Then the same holds for $\cC_y$ with any $y\in [y_1, y_2]$.}

\medskip
\noindent
{\bf Proof.}
Notice from the definition of $\cC_y$ that if  $\xi \in \cC_{y_1}$ and $\xi'\in \cC_{y_2}$, then
\be
(t \xi + (1-t) \xi') \in \cC_{t y_1 + (1-t) y_2}
\ee
holds for all $0\leq t \leq 1$.
Then apply this to such $\xi \in \cC_{y_1}$ and $\xi'\in \cC_{y_2}$ for which $\xi_n = \xi'_n=0$,
which exist since $\cC_y$ is stable under cyclic permutations of the components
of its elements.
\qed

\medskip
\noindent
{\bf Lemma 15.}  \emph{Choose  $0<y < \pi$ of the form
\be
y=\frac{\pi p}{n}  -\frac{\pi}{nq}
\quad\hbox{or}\quad
y=\frac{p\pi}{n} +\frac{\pi}{(n-q)n}
 \ee
with some $p$ and $q$  appearing in Theorem 12.
Then $\cC_y \cap \partial \cA \neq \emptyset$.}

\medskip
\noindent
{\bf Proof.} Following the proof of Proposition 11, one can show that in these cases
$\cC_y$ equals the simplex  $\cB(p,y)$, whose vertices now lie in $\partial \cA$.
Incidentally, these $y$ values do not satisfy (\ref{T1}). \qed

\medskip
\noindent
{\bf Lemma 16.} \emph{Suppose that $1<p <(n-1)$ satisfies $\gcd(n,p) \neq 1$. Then there exists $\varepsilon > 0$ such
that for any $y\in (p\pi/n, p\pi/n + \varepsilon)$ and for any
$y\in (p\pi/n - \varepsilon, p\pi/n)$ one has $\cC_y \cap \partial \cA\neq \emptyset$.}

\medskip
\noindent
{\bf Proof.}
For definiteness, consider the case $p\pi/n < y < (p+1) \pi/n$, when
\be
\cC_y = \cB(p,y) \cap \cB(p+1,y).
\ee
Then write $\ell := \gcd(n,p)$ and define the point $x$ by
\be
 x = (a_1,\cdots,a_\ell,a_1,\cdots,a_\ell,\cdots\cdots,a_1,\cdots,a_{\ell-1},0)
\ee
where
\be
 a_1 = \cdots = a_{\ell -1} = \frac{\frac{\ell}{p} y - a_\ell}{\ell - 1}
 \quad\hbox{and}\quad
  a_{\ell} = \frac{n}{p}y - \pi.
\ee
It is easily verified that $x \in E$. To see that $x \in \cA$, we need to
show that all $a_i \ge 0$. The fact that $a_\ell >0$ follows directly from $p\pi/n < y$.
For $a_i \ge 0$
for $1 \le i < \ell$, we need to have $y \le \frac{p\pi}{n-\ell}$, which is ensured
by a suitable choice of $\varepsilon$.

Since $a_1 + \cdots + a_\ell = \frac{\ell}{p}y$, it is readily checked that
$x \in \cB(p,y)$. To see that $x \in \cB(p+1,y)$, we argue as follows. A
cyclic permutation of the sum
\be
x_1 + \cdots + x_{p+1}
\ee
either contains the term $x_n$ or it does not. In the latter case, the sum
is clearly greater than $y$, since it contains all values $a_1,\cdots,a_\ell$ at least $\frac{p}{\ell}$ times.
In the former case, its value will be equal to
\be
y - a_{\ell} + a_i
\ee
for some $1 \le i < \ell$. So it is sufficient if
\be
a_i > a_\ell,
\ee
which can be ensured by possibly replacing $\varepsilon$ by a smaller value.

Now the proof is complete for $y\in (p\pi/n, p\pi/n + \varepsilon)$. The case
$y\in ( p\pi/n - \varepsilon, p\pi/n)$ can be handled in an analogous manner.
\qed

 \medskip
\noindent
{\bf Proof of Theorem 13.}
Suppose that $0<y< \pi$ subject to (\ref{T1})
does not belong to an open interval of the form (\ref{cpcase}).
(This excludes $n=2$ and $n=3$.)
Then, as is readily seen from Lemma 15 and Lemma 16, we can find $y_1$ and $y_2$ and integer $1 < k < (n-1)$ such that
$k \pi/n < y_1 < y <y_2 < (k+1)\pi/n$ and both $\cC_{y_1}$ and $\cC_{y_2}$ contain points of $\partial \cA$.
By using this and the fact that under (\ref{T1}) $\cA_y=\cC_y$, the required
statement results from Lemma 14.
\qed
\medskip

We end this section by a few remarks and questions.
We saw that the coupling parameters of the type (i) cases
are the generic $0<y<\pi$ values in the open intervals of the form
\be
(a_{p,n}\pi , b_{p,n}\pi) \quad \hbox{with}\quad
a_{p,n}= \frac{p}{n} - \frac{1}{nq}= \frac{m_p}{q},\quad
b_{p,n}= \frac{p}{n} + \frac{1}{n(n-q)} = \frac{p-m_p}{n-q},
\label{CPcase}\ee
where $p = 1,\ldots, (n-1)$, $\gcd(n,p)=1$ and $p q= m_p n + 1$.
These intervals enjoy the relation
\be
a_{n-p,n} = 1 - b_{p,n},
\quad
b_{n-p,n} = 1 - a_{p,n}.
\ee
It seems to be indicated by computer calculations that
every $y \neq p \pi/n$ from the interval
(\ref{CPcase}) satisfies (\ref{T1}), but we have not proved this.
In the type (i) cases the reduced phase space is $\cp$ carrying a multiple
of the Fubini-Study structure, but the constant involved was so far calculated
only when $p=1$ or $p=(n-1)$. See Section 4.

We observe that $ a_p \geq \frac{p-1}{n}$ having equality only for $p=1$, and
$b_p \leq \frac{p+1}{n}$ having equality only for $p=(n-1)$.
It is also not difficult to check that if $\gcd(n,p) =1$ and $\gcd(n,p+1) =1$ both hold
for some $1\leq p \leq (n-2)$, then
\be
b_{p,n} < a_{p+1,n}
\ee
except for $p=k$, $n=(2k+1)$ when
$b_{k,2k+1} = a_{k+1, 2k+1} = \frac{1}{2}$.
As a consequence, there exist $y$ values associated with type (ii)
reductions of $\SU(n)\times \SU(n)$  in every interval
$(\frac{j}{n}\pi, \frac{j+1}{n}\pi)$
for $j=1,\ldots, n-2$,
except for the interval $(\frac{k}{n}\pi, \frac{k+1}{n}\pi)$ if $n=2k+1$.
In particular,
type (ii) cases exist for every $n$ except for $n=2$ and $n=3$.
Taking $\SU(2k+1)$ with $k\geq 2$,
\be
  b_{1, 2k+1} \pi =\pi/(2k) < y < \pi/(k+1)= a_{2,2k+1}\pi
 \ee
  yields examples of type (ii) cases.
For $SU(2k)$ with $k\geq 2$,
$\pi/(2k-1) < y < \pi/k$ gives examples of type (ii) cases.

We have calculated the vertices
and faces of the  3-dimensional
``type (ii) convex polytope''
$\cA_y$ corresponding to $n=4$ and $\pi/3 < y < \pi/2$.
The vertices turned out to be  the cyclic permutations of the points
\be
R(1) := (y, \pi - 2y, 3y - \pi, \pi - 2y)
\quad\hbox{and}\quad
I(1):= (y, \pi - 2y, y, 0).
\label{E5}\ee
To describe the faces, let us  write $R(i)$ $(i=1,\ldots, 4)$
for the cyclic permutation $\sigma^{i-1}(R(1))$ of $R(1)$
using (\ref{cyclic}), and define $I(i)$ similarly.
Explicit inspection shows that $\cA_y$ possesses 4 triangular and 4 rectangular faces.
One particular triangular face is
incident with the vertices $R(1)$, $I(1)$ and $I(3)$,
and one  rectangular face
is incident with the vertices $R(2)$, $R(3)$, $I(3)$ and $I(4)$.
Then one can check that $I(1)$ is incident with two triangular faces and two rectangular faces. In three dimensions,
this means that $I(1)$ is incident with four edges. This implies that our 3-dimensional polytope $\cA_y$
is not a Delzant polytope,  since it is known \cite{De,Can}  that all vertices of the $n$-dimensional
Delzant polytopes are incident with precisely $n$ edges.
Of course it is not a surprise that $\cA_y$ is not a Delzant polytope, because
we do not obtain a toric structure in the type (ii) cases.
Interestingly, as follows from Corollary 4 in Section 2.2,
the regular vertices $R(i)$ correspond to fixed points of the $\beta$-generated torus action
on $\hat\beta^{-1}(\cA_y^\reg)$.
Concerning the interpretation of the irregular vertices $I(i)$,
we know from Appendix A that the position variables provided by $\hat\beta$ are not
differentiable at the locus $\hat\beta^{-1}(I(i))$, and the Hamiltonian vector fields
of the smooth reduced class functions
depending on $B$ from $[(A,B)]\in P(\mu_0(y))$ can span at most 2-dimensional
spaces at the points of $\hat\beta^{-1}(I(i))$, while generically they span 3-dimensional
subspaces of the tangent space.
Further details of this example, and the type (ii) systems in general, will be  studied
elsewhere.

\section{On new examples of type (i) cases}
\setcounter{equation}{0}

In the light of Theorem 12,
the standard compact RS systems associated with the coupling parameter $0< y < \pi/n$
represent examples of type (i) cases.
We have found new type (i) cases for which
the coupling parameter $y$ belongs to the interval
(\ref{cpcase}) for any $1\leq p\leq (n-1)$ with $\gcd(n,p)=1$.
(The cases associated with $p$ and $(n-p)$ are
essentially the same since $P(\mu_0(y))$ and $P(\mu_0(y)^{-1})$
are related by complex conjugation on the double.)
The goal of this section
is to elaborate certain details of new type (i) examples and explain
in what sense the corresponding compact RS systems  are different from the standard ones.
Specifically, we shall focus on the range of $y$ that lies on the right-side of $\pi/n$  in (\ref{cpcase})
for $p=q=1$,
i.e., we suppose that
\be
\frac{\pi}{n} < y < \frac{\pi}{(n-1)},
\qquad
n\geq 3.
\label{N1}\ee
By Proposition 11, the $\beta$-image
$\cA_y$ of the constraint surface is then given by
\be
\cA_y= \{ \xi \in \cA\,\vert\,\xi_\ell \leq y,\qquad \forall \ell=1,\ldots, n\}.
\label{NA}\ee
The  vertices of this simplex are $\xi(j)$ $(j=1,\ldots, n)$  having the components
\be
\xi(j)_\ell = y (1- \delta_{\ell, j}) + (\pi - (n-1) y) \delta_{\ell,j},
\qquad
j, \ell = 1,\ldots, n.
\label{N8}\ee
Since $\cA_y\subset \cA^\reg$,
the reduced phase space $P(\mu_0(y))$ is a Hamiltonian toric manifold
under the $\bT^{n-1}$-action generated by the  moment map
$\hat\beta^\lambda=\lambda \hat \beta$.
Thus one knows from the Delzant theorem \cite{De,Can} that $(P(\mu_0(y)), \omega_\red, \hat \beta^\lambda)$
is equivalent to $\cp$ equipped with the toric structure possessing
the same moment polytope $\lambda\cA_y$. We next describe the equivalence explicitly.
For definiteness, in what follows we assume that the overall parameter $\lambda$ in (\ref{I4}) is
positive.

Let us realize $\cp$ as a symplectic reduction of $\bC^n$ equipped with the symplectic form
$\Omega_{\bC^n} := \ri \sum_{k=1}^n \mathrm{d} \bar u_k \wedge \mathrm{d} u_k$.
This can be achieved by fixing the moment map
$\chi(u):= \sum_{k=1}^n \vert u_k \vert^2$
that generates the natural $U(1)$ action on $\bC^n$ (whereby $u\in \bC^n$ is mapped to $e^{\ri \gamma }u$).
Indeed, by applying the constraint
\be
\chi(u) = \chi_0 := \lambda(ny - \pi),
\qquad (\lambda>0),
\label{N10}\ee
the corresponding reduced phase space $\chi^{-1}(\chi_0)/U(1)$ turns out to be
\be
(\cp, \chi_0 \omfs),
\label{N11}\ee
where $\omfs$ is the standard Fubini-Study symplectic form.
Realizing any point of $\cp$ as an equivalence class $[u]= (u_1: u_2: \cdots : u_n)$ of some
$u \in \chi^{-1}(\chi_0)$,  we introduce the smooth functions $\cJ_k$ on $\cp$ by the definition
\be
\cJ_k([u]):= - \vert u_k \vert^2 + \lambda y,
\qquad
k=1,\ldots, n.
\label{N12}\ee
The definition ensures that
\be
\sum_{k=1}^n \cJ_k/\lambda = \pi
\quad\hbox{and}\quad
 \pi - (n-1)y  \leq \cJ_k/\lambda  \leq y, \qquad \forall k=1,\ldots, n.
 \label{N13}\ee
 The linearly independent functions $\cJ_k$ ($k=1,\ldots, n-1$) define
 the components of the moment map of a Hamiltonian action of $\bT^{n-1}$.
 This is the ``rotational action'' for which
 \be
 \tau = (\tau_1,\ldots, \tau_{n-1}) = (e^{\ri \theta_1}, \ldots, e^{\ri \theta_{n-1}})
\label{N14}\ee
operates by the map
\be
\cR_{\tau}: [(u_1, \ldots, u_{n-1}, u_n)] \mapsto [(\bar \tau_1 u_1,\ldots, \bar \tau_{n-1} u_{n-1}, u_n)],
\label{N15}\ee
i.e.,  by the Hamiltonian flow of $(\cJ_1,\ldots, \cJ_{n-1})$ at the ``time-parameters''
$(\theta_1,\ldots, \theta_{n-1})$.

The constants and the signs were purposefully chosen in the above definitions in such
a way that the image of the above toric moment map $\cJ$, where for convenience we
include in $\cJ$ the last component $\cJ_n = \lambda \pi - \sum_{k=1}^{n-1} \cJ_k$, is the same
polytope $\lambda \cA_{y}$ (\ref{NA}) that belongs to the $\beta$-generated $\bT^{n-1}$-action on
$P(\mu_0(y))$.
The vertices of the polytope correspond to the special points of $\cp$ where only one of the
homogeneous coordinates $(u_1,\ldots, u_n)$ is non-zero.

The Delzant theorem \cite{De,Can}  guarantees the existence of a diffeomorphism
\be
f_\beta: \cp \to P(\mu_0(y))
\label{N16}\ee
having the properties
\be
f_\beta^*(\omega_\red)= \chi_0 \omfs,
\qquad
f_\beta^* (\hat \beta^\lambda) = \cJ.
\label{N17}\ee
Such a map, called a ``Delzant symplectomorphism'', is essentially unique \cite{Pin}, that is,
it is unique up to the obvious possibility to compose it with the time-one flows
of arbitrary such Hamiltonians that can be expressed as functions
of the corresponding toric moment maps.

In order to construct  $f_\beta$, note that in the case under inspection Theorem 6
yields a symplectomorphism between
\be
(\cA_{y  }^+ \times \bT^{n-1}, \lambda \sum_{k=1}^{n-1} \mathrm{d}\theta_k \wedge \mathrm{d}\xi_k),
\label{N18}\ee
where $\cA_{ y }^+$ is the interior of $\cA_y$ in (\ref{NA}), and the dense open
submanifold $\hat\beta^{-1}(\cA_y^+) \subset P(\mu_0(y))$.
 Then introduce the map $\cE$ from the same domain (\ref{N18})  onto the dense open submanifold
  $\cp_0\subset \cp$ where none of the homogeneous coordinates vanish by setting
 \be
 \cE(\xi, \tau):= [\sqrt{\lambda} (\bar \tau_1 \sqrt{y - \xi_1} ,\ldots, \bar \tau_{n-1}\sqrt{y- \xi_{n-1}},
 \sqrt{y- \xi_n} )].
 \label{N19}\ee
 It is is easy to check that $\cE^*(\chi_0 \omfs)= \lambda \sum_{k=1}^{n-1} \mathrm{d}\theta_k \wedge \mathrm{d} \xi_k$ holds.

The composition of the above parametrizations of $\cp_0\subset \cp$ and $\hat\beta^{-1}(\cA_y^+)\subset P(\mu_0(y))$ by
$\cA_{y }^+ \times \bT_{n-1}$ gives rise to a symplectomorphism between $\cp_0$ and
$\hat\beta^{-1}(\cA_y^+)$,
which admits a global extension.
This is the content of the following theorem,  whose
proof is omitted since it is very similar to that of Theorem 5 in \cite{FK}.

\medskip
\noindent
{\bf Theorem 17.} \emph{The symplectomorphism $f_0: \cp_0 \to \hat\beta^{-1}(\cA_y^+)$ defined by
\be
f_0: \cE(\xi,\tau) \mapsto
\left[\left( g_y(\xi)^{-1}\cL_y^\loc(\xi,\tau)g_y(\xi), g_y(\xi)^{-1}\delta(\xi) g_y(\xi)\right)\right],
\qquad \forall (\xi,\tau)\in \cA^+_{y} \times \bT^{n-1},
\label{N20}\ee
where $[(A,B)]$ denotes the gauge orbit through $(A,B) \in \mu^{-1}(\mu_0(y))$, extends to a
global Delzant symplectomorphism $f_\beta$ verifying the properties (\ref{N17}).
}
\medskip

One of the key ingredients of the proof of Theorem 17 is to show that
after a suitable gauge transformation
the local Lax matrix $\cL^{\mathrm{loc}}_y$ (\ref{T45}) admits a smooth extensions from $\cp_0$ to $\cp$.
In fact, there exists a
unique function $\cL^y \in C^\infty(\cp, \mathrm{SU}(n))$ that satisfies
the  identity
\be
(\cL^y\circ \cE)(\xi,\tau) = \Delta(\tau)^{-1}
\cL^{\mathrm{loc}}_y(\xi, \tau)\Delta(\tau)
\quad\hbox{with}\quad \Delta(\tau):= \operatorname{diag}(\tau_1, \ldots, \tau_{n-1},1).
\label{N21}\ee
The function $\cL^y$ is called the global Lax matrix of the associated compact RS system.
Using the identification of the reduced phase space $P(\mu_0(y))$ with $\cp$ by the map $f_\beta$,
the compact RS system resulting from the reduction can be characterized by
the following properties:

\begin{enumerate}
\item{The global extension $H_y$ of principal RS Hamiltonian (\ref{T51}) transferred by
$f_0$ (\ref{N20}) to $\cp_0$ is given by
the real part of the trace of the global Lax matrix $\cL^y$, whose smooth class functions
generate an Abelian Poisson algebra on $(\cp, \chi_0 \omfs)$.}
\item{The functions $\cJ_k/\lambda = \hat \beta_k \circ f_\beta$
give globally smooth extension of the position variables $\xi_k$ of the local
RS system living on $\cA_y^+ \times \bT^{n-1}\simeq \cp_0$.}
\item{The functions
$\lambda \Xi_k \circ \cL^y = \lambda \hat \alpha_k \circ f_\beta$
define  globally smooth action variables for the compact RS system.}
\end{enumerate}
In conclusion,  the outcome of the reduction in the case (\ref{N1}) is
the compact RS system encoded by the triple $(\cp, \chi_0 \omfs, \cL^y)$ and
the above mentioned Abelian Poisson algebras of distinguished observables.

In the rest of this section,
we wish to compare the compact RS system that we just constructed
using the parameter $y$ subject to (\ref{N1}) to the original
compact system of Ruijsenaars \cite{RIMS95} having
the parameter $y$ in the range (\ref{I2}).
The physical interpretation of these systems is based on the ``principal local
Hamiltonian'' (\ref{T51}).
This Hamiltonian has the same form in all cases,
but different parameters $y$ appear
in it and the domain where the position variable $\xi$ is allowed to vary also depends on $y$.
Any two systems associated with different parameters are different in this basic sense.

We now further clarify the relation between the two systems by presenting them in terms
of the same coordinate system on $\cp_0$.
To elaborate this, let us denote all objects pertaining to the
``old case'' (\ref{I2})  by ``primed'' letters,
and also take the parameters positive.
Thus in the old case the reduced phase space is $\cp$ equipped with the symplectic
form
\be
\lambda' (\pi - n y') \omega_{\mathrm{FS}}
\quad\hbox{with}\quad 0 < y' < \pi/n.
\label{N22}\ee
The dense open submanifold of $\cp$ where none of the homogeneous coordinates vanish
is then parameterized  by the domain $\cA_{y'}^+ \times \bT^{n-1}$, where
$\cA^+_{y'}$ is  the Weyl alcove with thick walls
(\ref{I10}).
Concretely, the element
\be
(\xi', e^{\ri \theta_1'},\ldots, e^{\ri \theta_{n-1}'})
\in \cA_{y'}^+ \times \bT^{n-1},
\qquad
( y' <\xi_k',
\quad  \sum_{k=1}^n \xi_k'=\pi),
\label{N23}\ee
corresponds to the equivalence class
\be
[\sqrt{\lambda'} (e^{\ri \theta_1'} \sqrt{\xi_1' - y'} ,
\ldots,
 e^{\ri \theta'_{n-1}} \sqrt{\xi_{n-1}' -y'},
 \sqrt{\xi_n' -y'} )] \in \cp.
\label{N24}\ee
In this parametrization the symplectic form (\ref{N22}) becomes
$\lambda' \sum_{k=1}^{n-1} \mathrm{d} \theta_k' \wedge \mathrm{d} \xi_k'$
and the principal Hamiltonian reads
\be
H_{y'}^\loc(\xi', \theta') =
\sum_{j=1}^n \cos( \theta_j' - \theta_{j-1}') \prod_{k=j+1}^{j+n-1}
\left\vert 1 - \frac{\sin^2 y'}{ \sin^2(\sum_{m=j}^{k-1}  \xi_m')} \right\vert^\frac{1}{2}.
\label{N25}\ee

Since otherwise the resulting systems are plainly non-equivalent,
let us require  that in the old and new cases the reduction equips $\cp$ with the same
symplectic form, which means that the respective parameters $( \lambda', y')$
and $(\lambda, y)$ enjoy the relation
\be
\lambda' (\pi - n y') = \lambda (n y - \pi),
\label{N26}\ee
where $y'$ varies according to (\ref{N22}) and $\pi/n< y < \pi/(n-1)$.
The variables $\xi_k', e^{\ri \theta_k'}$ and $\xi_k, e^{\ri \theta_k}$
represent two coordinate systems on the same open dense submanifold $\cp_0 \subset\cp$, and thus
there is a unique relation between them. By comparing (\ref{N24}) and (\ref{N19}) under the assumption (\ref{N26}),
we find that the transformation between the coordinate systems is governed by the equations
\be
\theta_k = -  \theta_k',
\qquad
 \lambda (\xi_k - y) = \lambda' (y'   - \xi_k').
 \label{N27}\ee
 If we now express the ``new Hamiltonian'' $H_y^\loc$ in the primed variables by
 substituting the above formulas into (\ref{T51}), then we obtain the function
 \be
 H_y^\loc (\xi', \theta') =
 -\sum_{j=1}^n \cos(\theta_j' - \theta_{j-1}') \prod_{k=j+1}^{j+n-1}
\left\vert 1 - \frac{\sin^2 y}{ \sin^2(c_{j,k} +  (\lambda'/\lambda) \sum_{m=j}^{k-1}
\xi'_m)} \right\vert^\frac{1}{2}
\label{N28}\ee
with $c_{j,k} = \frac{\lambda'}{\lambda} (y+ y') (j-k)$.
It is clear that  when viewed as functions of the same coordinates on $\cp_0$
the  Hamiltonians  $H_{y'}^\loc(\xi', \theta')$ (\ref{N25}) and
$H_y^\loc(\xi', \theta')$ (\ref{N28}) are different.
Since their local restrictions are different,
$H_{y'}$ and $H_y$ are different functions on the full phase space $\cp$.
This holds even in those special cases for which the relations
$(\pi - n y') = (n y - \pi)$ and $\lambda' = \lambda$ are satisfied.
The conclusion is independent from having the overall minus sign in (\ref{N28}),
which comes from $s$ in (\ref{T51}) and could be dropped by change
of conventions or by suitable shifts of the variables $\theta_k'$.

To gain yet another perspective on the comparison,
note that we can express $H_y$ in terms its action variables
$I_k:=\lambda \hat\alpha_k$
and also express $H_{y'}$ in terms its action variables
$I'_k:= \lambda' \hat\alpha_k'$.
By using that $\hat \alpha$ and $\hat \beta$ have the same images due to (\ref{T39}), the Delzant theorem
guarantees the existence of a symplectomorphism that converts the respective action variables
into each other according to the relation
\be
\lambda (\hat \alpha_k -  y) \Longleftrightarrow \lambda' (y' - \hat \alpha_k').
\label{N29}\ee
This is fully analogous to the second equality in (\ref{N27}), where $\lambda\xi_k$ and $\lambda'\xi'_k$
are just the values taken by the toric moment maps $\lambda \hat \beta$ and $\lambda' \hat\beta'$.
The definition of the
function $\alpha$ (\ref{T4}) implies (by equation (\ref{fund}) in Appendix A) that
for $(A,B) \in \mu^{-1}(\mu_0(y))$ one has
$A \sim \exp(-2\ri \sum_{k=1}^{n-1} \hat \alpha_k \Lambda_k)$,
where $\sim$ means conjugation and we used the $n\times n$
matrices $\Lambda_k= \sum_{j=1}^k E_{j,j} - \frac{k}{n}\1_n$.
Then it is readily seen from the formulas
\be
H_y= \Re\tr \bigl(\exp(-2\ri \sum_{k=1}^{n-1} \hat \alpha_k \Lambda_k)\bigr)
\quad\hbox{and}\quad
H_{y'}= \Re\tr \bigl(\exp(-2\ri \sum_{k=1}^{n-1} \hat \alpha_k' \Lambda_k)\bigr)
\label{N30}\ee
that $H_y$ is \emph{not} converted into $H_{y'}$ by the symplectomorphism
that obeys (\ref{N29}). In other words, if we convert the action variables of the
unprimed system into the action variables of the primed system according
to (\ref{N29}), then $H_y$ and $H_{y'}$ become different functions of the
primed action variables $I'_k$.

The foregoing discussion can be informally summarized as follows:
``The systems associated with different parameters are at the first sight
obviously different, and this impression persists after closer inspection, too.''
It might be also possible to prove the non-existence of any symplectomorphism
of $\cp$ that would convert $H_y$ into $H_{y'}$ under the condition (\ref{N26}),
but we do not have such a proof.
The above arguments convinced us that no such symplectomorphism
exists if one requires it to have further natural properties, i.e.,
that it should map either particle positions into particle positions or action variables
into action variables.

\section{Conclusion}

In this paper we derived new compact forms of the trigonometric RS system
by reducing the quasi-Hamiltonian double of $G=\SU(n)$ at the moment map value
$\mu_0(y)$ (\ref{I1}) with generic angle parameter $y$.
These systems were previously considered in \cite{RIMS95,FK} under the restriction
$0<y < \pi/n$. We have shown that the reduction always yields a Liouville integrable system whose
 leading Hamiltonian  has the RS form  (\ref{I14}) on a dense open submanifold
 of the compact reduced phase space.
Different moment map values (with $0<y< \pi/2$) correspond to inequivalent
many-body systems in general.
It turned out that
two drastically different types of cases occur, which we termed type (i) and type (ii).

In the type (i) cases the reduced phase space $P(\mu_0(y))$ is a Hamiltonian toric manifold since it
inherits globally smooth action and position variables from the double.
Our main result (given by Theorems 12 and 13 in Section 3) is that
we found all $y$ values associated with type (i) cases,
and also found that the pertinent toric moment polytope is always a simplex.
This implies the existence of an equivariant symplectomorphism between the
reduced phase space $P(\mu_0(y))$ and the complex projective space equipped with a
multiple of its standard symplectic structure, which we detailed
for the particular type (i) cases having coupling parameter $\pi/n <y < \pi/(n-1)$.

In the type (ii) cases
the action and position variables lose their differentiability
on a nowhere dense subset of $P(\mu_0(y))$.
The existence of such cases is an unexpected new result.
The properties of the corresponding compact RS systems
 should be further explored in the future.

We worked at the classical level, but the quantum mechanics of our
systems should be also investigated.
It is more or less clear how to perform such investigation in the type (i) cases,
since there exist general results on the quantization of Hamiltonian toric manifolds \cite{Ham}
and also a detailed study \cite{vDV}  of the quantum mechanics of the standard compact  RS systems
 belonging to the range $0<y <\pi/n$.
In the type (ii) cases no previous studies exist.

Finally, it is worth stressing that
the compact RS systems (both type (i) and type (ii)) that we dealt with are self-dual
in the sense that there exists a symplectomorphism of order $4$ on their
phase space exchanging the position and action variables.
In the same way as explained in \cite{FK},
the self-duality map descends from the natural action of the modular $\SL(2,\bZ)$ group on the double, which
provides a finite dimensional model for describing  the moduli
spaces of flat $\SU(n)$ connections on the one-holed torus \cite{AMM}.
It should be possible to construct a corresponding
quantum mechanical representation of the $\SL(2,\bZ)$ group in the compact RS systems.
General arguments based  on Chern-Simons field theories \cite{Wi} and on Hecke algebras \cite{Cher}
indicate the existence of such $\SL(2,\bZ)$ representation, but its construction
in sufficiently concrete terms was, as far as we know,  not addressed before even in the standard case
\cite{vDV}.

\renewcommand{\theequation}{\Alph{section}.\arabic{equation}}
\renewcommand{\thesection}{\Alph{section}}
\setcounter{section}{0}

\section{Some properties of class functions of $G$}
\setcounter{equation}{0}

In this appendix we briefly survey relevant properties of the real class functions of
$G:=\SU(n)$.
We first show that  the derivatives of globally smooth class functions span an
$(n-1)$-dimensional space at all regular points,
but a smaller dimensional subspace at singular points.
Then we explain that the class functions $\Xi_k$ that we defined in (\ref{T3}) are not globally
smooth. They are smooth when restricted to $G_\reg$ and only continuous at
$G_\sing$.
These results are well known in Lie theory,
and are described here to make our text essentially self-contained.

To begin,  let us remark that at any $g\in G$ the $\cG$-valued derivative
$\nabla h(g)$ of $h\in C^\infty(G)^G$ (which is the translate of the
usual exterior derivative to the unit element)
belongs to the center of the Lie algebra
of the stabilizer subgroup $G_g$ of $g$ with respect to conjugation.
This is a consequence
of the equivariance property $\nabla h \in C^\infty(G,\cG)^G$.
At regular $g$, $G_g$ is Abelian of dimension $(n-1)$, while at non-regular $g$
the dimension of the center of the Lie algebra of $G_g$ is smaller than $(n-1)$.
Thus it follows that at $g\in G_\sing := G\setminus G_\reg$ the dimension of the
span of the derivatives of the $C^\infty$ class functions drops; it
becomes zero at the center of $G$.
Via our reduction, the
smooth class function applied to $A$ in $(A,B)\in G\times G$
 descend to the globally smooth
 principal Hamiltonian of the compact RS systems and its commuting family.
The dimension of the span of the derivatives of the functions concerned
cannot increase through the reduction, which involves projections.
(It can actually decrease, as is exemplified by the vertices of the Delzant polytope (\ref{N8}),
where the Hamiltonian vectors fields of all reduced ``smooth class functions of $B$'' vanish.)
The message is that interesting special phenomena in the behaviour
of the Hamiltonian flows can be
expected at the points of the reduced phase space that come from gauge orbits
for which $A$ or $B$ in $(A,B) \in \mu^{-1}(\mu_0(y))$ belongs to $G_\sing$.

Next, let us focus on the ``spectral functions'' $\Xi_k$ (\ref{T3}) that were
crucial for our considerations.
These were defined using the formula (\ref{I8}), which can be recast in the equivalent form
\be
\delta(\xi) = \exp\left(-2 \ri \sum_{k=1}^{n-1} \xi_k \Lambda_k\right),
\label{fund}\ee
where the diagonal matrices
$\Lambda_k =\sum_{j=1}^k E_{j,j} - \frac{k}{n} \1_n$
realize the fundamental weights of $\su(n)$ in the standard manner.
Every conjugacy class of $G$ admits a representative of the form
 $\delta(\xi)$ for a unique $\xi \in \cA$.
 Thus  formula (\ref{fund}) yields a
 one-to-one correspondence between the elements of the alcove $\cA$ (\ref{I7}) and
the conjugacy classes  of $G$.
This correspondence is known to be a \emph{homeomorphism} \cite{DT}
with respect to the topology on the set of conjugacy classes inherited from the group
and the topology on the alcove $\cA$ inherited from its embedding in
 $\bR^n$  (or in the Lie algebra of the maximal torus).
 Hence our spectral functions $\Xi_k$ are continuous functions on $G$.
 It is also well known that the mapping
\be
\cA^\reg \times (G/\bT^{n-1}) \to G_\reg
\quad\hbox{defined by}\quad
(\xi, \gamma \bT^{n-1}) \mapsto  \gamma \delta(\xi) \gamma^{-1},
\ee
where $\cA^\reg$ is the interior of the alcove $\cA$,
is an analytic diffeomorphism of real analytic manifolds.
In particular,  the spectral functions
are real analytic  (and thus also smooth) functions on $G_\reg$.
They encode the $\cA$-component of the analytic inverse of the above map.

The  parametrization by the representatives in (\ref{fund}) is a special case
of the parametrization of the conjugacy classes by a fundamental
domain of the affine Weyl group, which works similarly for any connected and simply
connected  simple compact Lie group \cite{DT}.

Finally, let us explain the non-differentiability of the spectral functions at the
singular locus $G_\sing$.
As an illustration,  consider the group $\SU(2)$ and parametrize the elements $\eta$
from a small neighbourhood of the identity in its maximal torus  as
\be
\eta(x) := \diag(e^{\ri x}, e^{-\ri x}),
\qquad
x\in (-\epsilon, \epsilon).
\ee
It is not hard to see
from the definition (\ref{T3}) that the first component of $\Xi:= \Xi^{\SU(2)}$ satisfies
\be
\Xi^{\SU(2)}_1(\eta(x)) = \vert x \vert
\ee
for small $x$. This function is not differentiable at $x=0$.

In order to  demonstrate that the spectral functions of $G=\SU(n)$ for $n>2$  are also not differentiable
at $G_\sing$,   suppose that the converse was true.
That is, suppose
that $\Xi^G$ is smooth at $g\in G_\sing$.
We show that this would imply the smoothness of $\Xi^{\SU(2)}$ at the
identity (contradicting what we have seen).
To do this, take $g\in G_\sing$  as a diagonal matrix in the normal form (\ref{fund}),
and assume that $\xi_i=0$ for some $1\leq i\leq (n-1)$, which means that $\delta_{i} = \delta_{i+1}$.
For simplicity, we also assume that all other components of $\xi$ are positive.
Then define the smooth map $F$  by
\be
F\colon \SU(2) \to \SU(n), \qquad
\eta \mapsto  \diag(\delta_1,\cdots,\delta_{i-1}, \eta \delta_i,\delta_{i+2},\cdots,\delta_n) ,
\ee
where the instance of $\diag$ should be read as a block-diagonal matrix.
It is easy to check that
\be
\Xi^{\SU(2)}_1(\eta(x)) = (\Xi^G_i \circ F)(\eta(x))
\ee
near the identity. Then, because $\Xi^{G}_i$ is smooth by assumption and because $F$ is smooth by definition,
so would be $\Xi^{\SU(2)}_1$. This contradictions shows that our assumption is false. In other words,
$\Xi^{G}$ is not smooth at $\delta(\xi)\in G_\sing$. Similar arguments can be applied
to demonstrate non-smoothness at arbitrary points of $G_\sing$.

The local properties of the spectral functions also follow
from classical results about the behavior of (ordered) eigenvalues
of matrices under multi-parameter analytic perturbations \cite{Kato}.

\section{Denseness properties}
\setcounter{equation}{0}

Our purpose is to show that $\hat \beta^{-1}(\cA_y^+)$, where
the local RS system lives according to Theorem 6, is a \emph{dense}  submanifold
of the reduced phase space. If (\ref{T2}) holds, this easily follows from the fact
that $\hat \beta^{-1}(\cA_y^+)$ is exactly the subset of principal orbit type for the $\beta$-generated
torus action on the Hamiltonian
toric manifold $P(\mu_0(y))$,
which is known to be dense. If (\ref{T2}) fails, however, we do not have a Hamiltonian toric
manifold structure on $P(\mu_0(y))$, necessitating a separate proof.

 We first demonstrate that the $\beta$-regular part of the
constraint surface is dense.

\medskip
\noindent
{\bf Proposition B.1.} \emph{For any $y$ in (\ref{T1}), the elements $(A,B) \in G \times G$
such that $\mu(A,B) = \mu_0(y)$ and
$B$ is regular form a dense subset of the solutions to
$\mu(A,B) = \mu_0(y)$.}

\medskip
\noindent
{\bf Proof.}
Recall the definition of the discriminant of a polynomial $f$:
\be
\Delta(f) := \prod_{i<j}(\lambda_i - \lambda_j)^2
\ee
for $f$ given by
\bea
f &=& (\lambda - \lambda_1)\cdots(\lambda - \lambda_n).
\eea
It is a classical result that $\Delta(f)$ is actually a polynomial in the coefficients of $f$.
It is clear that
$\Delta(f)$ is zero exactly when $f$ has a double zero.

We know that $\mu^{-1}(\mu_0(y))$ is a connected, regular submanifold
of $G\times G$. In fact, since the moment map
constraint is a set of polynomial equations, we also know that $\mu^{-1}(\mu_0(y))$
inherits  an \emph{analytic}\footnote{Our use of `analytic' in this appendix always means
`real analytic'.}
manifold structure from $G\times G$.
Thus the matrix elements of $A$ and $B$ are analytic functions on it.

We define the complex function $\phi$ on $ \mu^{-1}(\mu_0(y))$ by
\be
\phi: (A,B) \mapsto  \Delta(\det(\lambda - B)).
\ee
It vanishes exactly when $B$ has a double eigenvalue.
By the above, $\phi$ is an analytic function on $\mu^{-1}(\mu_0(y))$.
If $\phi^{-1}(\{0\})$ has non-empty interior, then $\phi$ must vanish
identically on $\mu^{-1}(\mu_0(y))$, since it is  an analytic connected manifold.
This proves that either the subset of $\mu^{-1}(\mu_0(y))$ for which $B$ is
non-regular has empty interior, or it
coincides with $\mu^{-1}(\mu_0(y))$.

We know, however, that there exists a solution $(A,B)$ to the moment map constraint for which
$\beta(A,B) = \xi^*$ with $\xi^*$ defined in (\ref{T25}).
Since every component of $\xi^*$ is positive, this $B$ is regular. This shows that $\phi$ does not vanish
identically, and thereby the proposition is proved.
\qed

\medskip\noindent
{\bf Corollary B.2.} \emph{For any $y$ in (\ref{T1}), $\mu^{-1}(\mu_0(y)) \cap (G_\reg \times G_\reg)$
is a dense open submanifold of the constraint surface $\mu^{-1}(\mu_0(y))$.}

\medskip\noindent
{\bf Proof.}
Proposition B.1 ensures that $\mu^{-1}(\mu_0(y)) \cap (G \times G_\reg)$ is a dense open
subset of $\mu^{-1}(\mu_0(y))$,
and $\mu^{-1}(\mu_0(y)) \cap (G_\reg \times G)$ clearly enjoys the same property.
The intersection of two dense open sets is again dense open.
\qed

Since the image of a dense set under a continuous surjective map is dense,
it follows from Proposition B.1 that the subsets given in the next line are \emph{dense}:
\be
\hat\beta^{-1}(\cA_y^\reg) \subset P(\mu_0(y))
\quad
\hbox{and}\quad
\cA_y^\reg \subset \cA_y.
\ee
We now wish to prove that analogous statements hold also for $\cA_y^+ \subset \cA_y^\reg$
defined in (\ref{T44}).
Our argument will be very similar to the proof of Proposition B.1.

\medskip
\noindent
{\bf Proposition B.3.}
\emph{The open submanifold $\beta^{-1}(\cA_y^+) \cap \mu^{-1}(\mu_0(y))$ of
the constraint surface is a dense subset of $\beta^{-1}(\cA_y^\reg) \cap \mu^{-1}(\mu_0(y))$.
}

\medskip
\noindent
{\bf Proof.}
Using (\ref{T15}),
define the real function $\psi$ on the analytic manifold
$\beta^{-1}(\cA_y^\reg)\cap \mu^{-1}(\mu_0(y))$ by the formula
\be
\psi: (A,B) \mapsto \prod_{\ell=1}^n z_\ell(\Xi(B),y).
\ee
Since $ \Xi: G_\reg \to \cA^\reg$ is an analytic map,  it
follows that $\psi$ is analytic.

Note that the submanifold $\beta^{-1}(\cA_y^+) \cap \mu^{-1}(\mu_0(y))$ is exactly
the subset of $\beta^{-1}(\cA_y^\reg) \cap \mu^{-1}(\mu_0(y))$ where $\psi$ takes non-zero values.
Suppose that it is not a dense subset. Then there exists a non-empty open subset of
$\beta^{-1}(\cA_y^\reg) \cap \mu^{-1}(\mu_0(y))$ on which $\psi$ vanishes identically. Because $\psi$ is
analytic, this implies that $\psi$ vanishes identically on an entire connected component $M$ of
$\beta^{-1}(\cA_y^\reg) \cap \mu^{-1}(\mu_0(y))$.

Let $\hat M$ be the connected component of $\hat \beta^{-1}(\cA_y^\reg)\subset P(\mu_0(y))$ corresponding to $M$,
and let $\hat M_0$ be the dense open subset of $\hat M$ containing the points of principal orbit
type for the $\beta$-generated $\bT^{n-1}$-action restricted to $\hat M$.
Since $\psi$ vanishes
on $M$, it follows from (the discussion following) Lemma 3 that the $\bT^{n-1}$-action on $\hat M$ has
orbits of dimension strictly smaller than $n-1$.
Moreover, it follows from the  theorem on principal orbit type (e.g.~\cite{DT}) that $\hat M_0$ is a locally trivial fibre bundle.
Suppose that the $\bT^{n-1}$ orbits in $\hat M_0$ are of dimension $r$.
Using that the $\bT^{n-1}$-action is generated by the moment map
$\hat \beta$ and is transitive on $\hat \beta^{-1}(x)$
for all $x\in \cA_y^\reg$,
we then see that
the restriction of the map $\hat \beta$ to $\hat M_0$
 induces a smooth one-to-one map of constant rank $r$ from the base of the bundle $\hat M_0$
 into  $\cA^\reg$.
 This would imply that the dimension of $\hat M_0$ equals $2r < 2(n-1)$,
 which contradicts $\hat M_0$ being an open submanifold of the reduced phase space of dimension $2(n-1)$.
  This  contradiction shows that
it is not possible for the connected component $M$ to be fully contained in the zero set of $\psi$. Therefore,
our assumption that the submanifold $\beta^{-1}(\cA_y^+) \cap \mu^{-1}(\mu_0(y))$ is not dense was false,
proving the proposition.
\qed

\medskip
\noindent
{\bf Corollary B.4.}  \emph{The following is a chain of dense open submanifolds of the reduced phase space:
\be
\hat\beta^{-1}(\cA_y^+) \subset \hat \beta^{-1}(\cA_y^\reg) \subset P(\mu_0(y)),
\ee
and $\cA_y^+ \subset \cA_y$ is a dense subset.
}
\medskip

Corollary B.4  shows that the local RS system of Theorem 6 always
lives on an open dense submanifold of the reduced phase space, which is what we wanted to
prove.

\bigskip
\bigskip
\noindent{\bf Acknowledgements.}
LF wishes to thank  Y. Karshon, J. Kincses and T.F. G\"orbe for correspondence and for discussions.
This work was supported in part by the Hungarian Scientific Research Fund (OTKA) under the grant
K 77400 and by the project T\'AMOP-4.2.2.A-11/1/KONYV-2012-0060
financed by the EU and co-financed by the European Social Fund. The work of TK was supported by the Utrecht University
program `Foundations of Science'.
This research was helped by computer explorations using the open-source mathematical software Sage \cite{SAGE}
and its interface to the software package Polymake \cite{Polymake} for calculating polytope properties.

\end{document}